\documentclass[11pt]{article}

\usepackage{amsmath,amssymb}

\def\tabaddress#1{{\small\it\begin{tabular}[t]{c}#1
\\[1.2ex]\end{tabular}}}

\font\fr=eufm10 scaled \magstep 1 
\newtheorem{teor}{Theorem}
\newtheorem{prop}{Proposition}
\newtheorem{corol}{Corollary}
\newtheorem{definition}{Definition}
\newtheorem{lem}{Lemma}
\newtheorem{remark}{Remark}
\def\beq{\begin{equation}}
\def\eeq{\end{equation}}
\def\bea{\begin{eqnarray}}
\def\eea{\end{eqnarray}}
\def\beann{\begin{eqnarray*}}
\def\eeann{\end{eqnarray*}}
\def\ben{\begin{enumerate}}
\def\een{\end{enumerate}}
\def\bit{\begin{itemize}}
\def\eit{\end{itemize}}
\def\derpar#1#2{\frac{\partial{#1}}{\partial{#2}}}

\def\feble#1{\mathrel{\mathop =\limits_{#1}}}

\def\moment#1#2#3{{#1}_{#2}, \ldots, {#1}_{#3}}

\def\qed{\ifvmode\removelastskip\fi
{\unskip\nobreak\hfil\penalty50\hbox{}\nobreak\hfil \hbox{\vrule
height1.2ex width1.2ex}\parfillskip=0pt \finalhyphendemerits=0
\par\smallskip}}
\def\vf{\mbox{\fr X}}
\def\df{{\mit\Omega}}
\def\Lag{{\cal L}}
\def\lag{\pounds}

\def\d{{\rm d}}

\def\Real{\mathbb{R}}

\def\inn{\mathop{i}\nolimits}
\def\Tan{{\rm T}}

\def\Lie{\mathop{\rm L}\nolimits}
\def\ls{(J^1\pi,\Omega_\Lag)}
\def\hsjpi{(J^1\pi^*,h)}
\def\hsmpi{({\cal M}\pi,\Omega,\alpha)}
\def\hsjpio{(J^1\pi^*,{\cal P},h_{\cal P})}
\def\hsmpio{({\cal M}\pi,\tilde{\cal P},\alpha_{\tilde{\cal P}})}

\def\Cinfty{{\rm C}^\infty}
\def\proof{( {\sl Proof} )\quad}

\title{EXTENDED HAMILTONIAN SYSTEMS IN MULTISYMPLECTIC FIELD THEORIES}
\author{\sc Arturo Echeverr\'\i a-Enr\'\i quez \\
 \tabaddress{Departamento de Matem\'atica Aplicada IV\\
  Edificio C-3, Campus Norte UPC.
  C/ Jordi Girona 1. E-08034 Barcelona. Spain}
   \\
{\sc Manuel de Le\'on\thanks{{\bf e}-{\it mail}: mdeleon@imaff.cfmacc.csic.es}}
 \\
 \tabaddress{Instituto de Matem\'aticas y F\'\i sica Fundamental, CSIC\\
   C/ Serrano 123. E-28006 Madrid. Spain}
   \\
{\sc Miguel C. Mu\~noz-Lecanda\thanks{{\bf e}-{\it mail}:
 matmcml@ma4.upc.edu}},
{\sc Narciso Rom\'an-Roy\thanks{{\bf e}-{\it mail}:
  nrr@ma4.upc.edu}}
\\
 \tabaddress{Departamento de Matem\'atica Aplicada IV\\
  Edificio C-3, Campus Norte UPC.
  C/ Jordi Girona 1. E-08034 Barcelona. Spain}}

\begin{document}
\maketitle
\thispagestyle{empty}
\setcounter{page}{0}

 \begin{abstract}
 We consider Hamiltonian systems in first-order multisymplectic field theories.
 We review the properties of Hamiltonian systems
 in the so-called {\sl restricted multimomentum bundle}, including
 the variational principle which leads to the Hamiltonian field equations.
 In an analogous way to how these systems are defined
 in the so-called {\sl extended} (symplectic) {\sl formulation}
 of non-autonomous mechanics, we introduce Hamiltonian systems
 in the {\sl extended multimomentum bundle}.
 The geometric properties of these systems are studied,
 the Hamiltonian equations are analyzed
 using integrable multivector fields,
 the corresponding variational principle is also stated, and
 the relation between the extended and the restricted Hamiltonian systems
 is established.
 All these properties are also adapted to 
 certain kinds of submanifolds of the
 multimomentum bundles in order to cover the case of
 almost-regular field theories.
 \end{abstract}


\bigskip
\bigskip

\vbox{\raggedleft AMS s.\,c.\,(2000):
 70S05, 55R10, 53C80.\\
PACS (1999): 02.40.Hw, 02.40.Vh, 11.10.Ef, 45.10.Na }\null

 \clearpage

 \tableofcontents

\clearpage

 \section{Introduction}

The Hamiltonian formalism of dynamical systems, and the study of the properties of
Hamiltonian dynamical systems in general, is a fruitful subject
in both applied mathematics and theoretical physics. From a
generic point of view, the characteristics of these kinds of systems
make them specially suitable for analyzing many of their properties;
for instance: symmetries and related topics such as the existence
of conservation laws and reduction, the integrability (including
numerical methods), and the possible quantization of the system,
which is based on the use of the Poisson bracket structure of this
formalism. Moreover, it is also important to point
out the existence of dynamical Hamiltonian systems which have no
Lagrangian counterpart (see an example in \cite{So-69}).

 From the geometrical viewpoint, many of the
characteristics of the autonomous Hamiltonian systems arise from
the existence of a ``natural'' geometric structure with which the phase
space of the system is endowed: the {\sl symplectic form}
(a closed, nondegenerated two-form), which allows the construction of
Poisson brackets. In this model, the dynamic information is
carried out by the {\sl Hamiltonian function}, which is not
coupled to the geometry.
This is not the case for non-autonomous Hamiltonian systems,
which have different geometric descriptions.
One of the most frequently used formulations for these systems
is in the framework of contact geometry,
which takes place in the restricted phase space $\Tan^*Q\times\Real$,
where $Q$ is the configuration manifold
(see \cite{EMR-91} and references therein).
Here, the physical information is given by the Hamiltonian function,
which allows us to construct the contact form in $\Tan^*Q\times\Real$.
However, a more appropriate description is
the symplectic or extended formulation of non-autonomous mechanics
\cite{GMS-97}, \cite{Ku-tdms}, \cite{MS-98}, \cite{Ra1}, \cite{St-2005}, which
is developed in the extended phase space $\Tan^*(Q\times\Real)$.
Now, the natural symplectic structure of $\Tan^*(Q\times\Real)$
and the physical information, given by the extended Hamiltonian function,
are decoupled and this provides us with a Hamiltonian description similar to the autonomous case.

When first-order field theories are considered, the usual way to
work is with the Lagrangian formalism \cite{AA-80}, \cite{BSF-88},
\cite{EMR-96}, \cite{GS-73}, \cite{Gc-73}, \cite{GMS-97b},
\cite{mod2}, \cite{Ma-2005}, \cite{Sa-89}, because their Hamiltonian description
presents different kinds of problems. First, several Hamiltonian
models can be stated, and the equivalence among them is not always
clear (see, for instance, \cite{Aw-92}, \cite{EMR-00},
\cite{Go-91b}, \cite{Gu-87}, \cite{HK-01}, \cite{Ka-98},
\cite{fam}, \cite{Sd-95}). Furthermore, there are equivalent Lagrangian models
with non-equivalent Hamiltonian descriptions \cite{Kr-87},
\cite{KS-01a}, \cite{KS-01b}. Among the different geometrical
descriptions to be considered for describing field theories, we
focus our attention on the multisymplectic models \cite{CM-2003},
\cite{GIMMSY-mm}, \cite{KS-75}, \cite{KT-79},  \cite{Ma-2004};
where the geometric background is in the realm of multisymplectic manifolds,
which are manifolds endowed with a closed and $1$-nondegenerate
$k$-form, with $k\geq 2$. In these models, this form plays a similar role to the
symplectic form in mechanics.

The main aim of this paper is to generalize the Hamiltonian
symplectic formulation of non-autonomous mechanics to
first-order multisymplectic field theories.
The motivation and basic features of this formulation are the following:
As is well known, there is no a canonical model
for Hamiltonian first-order field theory.
Hence the first problem to be considered is the choice of a suitable
multimomentum bundle to develop the formalism.
The most frequently used choice is to take the so-called {\sl restricted
multimomentum bundle}, denoted by $J^1\pi^*$; 
that is analogous to $\Tan^*Q\times\Real$ in the mechanical case.
The Hamiltonian formalism in $J^1\pi^*$ has been
extensively studied \cite{CCI-91}, \cite{EMR-99b}, \cite{LMM-96},
\cite{MS-99}. Nevertheless, this bundle does not have a canonical
multisymplectic form and the physical information, given by a
Hamiltonian section, is used to obtain the geometric structure.
This is a problem when other aspects of Hamiltonian field theories are
considered, such as: the definition of Poisson brackets,
the notion of integrable system,
the problem of reduction by symmetries,
and the quantization procedure.
An attempt to overcome these difficulties is to work in a greater
dimensional manifold, the so-called {\sl extended multimomentum
bundle}, denoted by ${\cal M}\pi$,
which is the analogous to the extended phase space $\Tan^*(Q\times\Real)$
of a non-autonomous mechanical system.
${\cal M}\pi$ has a canonical multisymplectic form,
since it is a vector subbundle of a multicotangent bundle.
In this manifold ${\cal M}\pi$, the physical information is given
by a closed one form, the Hamiltonian form. Then 
Hamiltonian systems can be introduced as in autonomous
mechanics, by using certain kinds of Hamiltonian multivector fields.
The resultant {\sl extended Hamiltonian formalism} is the
generalization to field theories of the extended formalism for
non-autonomous mechanical systems \cite{Ku-tdms}, \cite{EMR-91} and, to our
knowledge, it was introduced for the first time in field theories
in \cite{PR-2002}.

The goal of our work is to carry out a deeper geometric study of these
kinds of systems.
The main results are the following: first, to every Hamiltonian
system in the extended multimomentum bundle, we can associate
in a natural way a class of equivalent Hamiltonian systems in the
restricted multimomentum bundle (Theorem \ref{related}),
and conversely (Proposition \ref{existuniq}).
The solutions to the field equations
in both models are also canonically related.
In addition, the field equations for these kinds of systems can be derived 
from an appropriate variational principle (Theorem \ref{exequics}),
which constitutes a first attempt to tackle variational principles
for field theories with non-holonomic constraints
(see \cite{VCLM-2005} for a geometrical setting of these theories).
Furthermore, the integral submanifolds of the
Hamiltonian $1$-form can be embedded into the extended multimomentum phase space
similar to the way in which the constant energy surfaces are
coisotropically embedded in $\Tan^*(Q\times\Real)$ in
non-autonomous mechanics (Proposition \ref{coiso}).
Finally, the case of {\sl non regular Hamiltonian systems} is considered and,
after a carefull definition of what an {\sl almost-regular} Hamiltonian system is,
the above results are adapted to this situation in a natural way.
We hope that all these results could be a standpoint
from which to study Poisson brackets, the quantization problem 
and also the reduction by symmetries of field theories
in further research works.

The paper is organized as follows: In Section \ref{mmb} we review
basic concepts and results, such as multivector fields and
connections, multisymplectic manifolds and Hamiltonian multivector
fields, and the restricted and extended multimomentum bundles with
their geometric structures. Section \ref{hs} is devoted to reviewing
the definition and characteristics of Hamiltonian systems in the
restricted multimomentum bundles; in particular, the definitions
of Hamiltonian sections and densities, the variational principle
which leads to Hamilton-De Donder-Weyl equations, and the use of
multivector fields for writing these equations in a more suitable geometric way.
Sections \ref{ehs} and \ref{arhs} contain the most relevant
material of the work. Thus,
Hamiltonian systems in the extended multimomentum bundle are
introduced in Section \ref{ehs}; in particular, their
geometric properties, their relation with those introduced in Section \ref{hs},
and the corresponding variational principle are studied.
In Section \ref{arhs} we adapt the above
definitions and results in order to consider Hamiltonian
systems which are not defined everywhere in the multimomentum bundles,
but in certain submanifolds of them: here,
these are the so-called {\sl almost regular Hamiltonian
systems}. Finally, as typical examples, in Section \ref{exs}, we review the
standard Hamiltonian formalism associated to a Lagrangian field
theory, both in the regular and singular (almost-regular) cases,
and the Hamiltonian formalisms of time-dependent dynamical systems
in the extended and restricted phase space, which are recovered as
a particular case of this theory.

All manifolds are real, paracompact, connected and $C^\infty$. All
 maps are $C^\infty$. Sum over crossed repeated indices is
 understood. Throughout this paper
 $\pi\colon E\to M$ will be a fiber bundle
 ($\dim\, M=m$, $\dim\, E=n+m$),
 where $M$ is an oriented manifold with volume form
 $\omega\in\df^m(M)$, and $(x^\nu,y^A)$
 (with $\nu = 1,\ldots,m$; $A= 1,\ldots,n$) will be
 natural local systems of coordinates in $E$
 adapted to the bundle, such that
 $\omega=\d x^1\wedge\ldots\wedge\d x^m\equiv\d^mx$.

 \section{Previous definitions and results}
 \protect\label{mmb}

\subsection{Multivector fields and connections}
\protect\label{mvfc}

(See \cite{EMR-98} for details).

Let ${\cal M}$ be a $n$-dimensional differentiable manifold.
Sections of $\Lambda^m(\Tan {\cal M})$ are called
$m$-{\sl multivector fields} in ${\cal M}$
(they are the contravariant skew-symmetric tensors of order $m$ in ${\cal M}$).
We will denote by $\vf^m ({\cal M})$ the set of
$m$-multivector fields in ${\cal M}$.

If ${\cal Y}\in\vf^m({\cal M})$, for every $p\in {\cal M}$,
 there exists an open neighbourhood
 $U_p\subset {\cal M}$ and $Y_1,\ldots ,Y_r\in\vf (U_p)$ such that
\(\displaystyle {\cal Y}\feble{U_p}\sum_{1\leq i_1<\ldots <i_m\leq r}
f^{i_1\ldots i_m}Y_{i_1}\wedge\ldots\wedge Y_{i_m}\),
with $f^{i_1\ldots i_m}\in\Cinfty (U_p)$ and $m\leq r\leq{\rm dim}\, {\cal M}$.
Then, ${\cal Y}\in\vf^m({\cal M})$ is said to be {\sl locally decomposable} if,
for every $p\in {\cal M}$, there exists an open neighbourhood 
$U_p\subset {\cal M}$
and $Y_1,\ldots ,Y_m\in\vf (U_p)$ such that
${\cal Y}\feble{U_p}Y_1\wedge\ldots\wedge Y_m$.

A non-vanishing $m$-multivector field ${\cal Y}\in\vf^m({\cal M})$ and
a $m$-dimensional distribution $D\subset\Tan {\cal M}$
are {\sl locally associated} if there exists a connected open set
$U\subseteq {\cal M}$ such that ${\cal Y}\vert_U$ 
is a section of $\Lambda^mD\vert_U$.
If ${\cal Y},{\cal Y}'\in\vf^m({\cal M})$ are non-vanishing multivector fields
locally associated with the same distribution $D$,
on the same connected open set $U$, then there exists a
non-vanishing function $f\in\Cinfty (U)$ such that
${\cal Y}'\feble{U}f{\cal Y}$. This fact defines an equivalence relation in the
set of non-vanishing $m$-multivector fields in ${\cal M}$, 
whose equivalence classes
will be denoted by $\{ {\cal Y}\}_U$. Then
there is a one-to-one correspondence between the set of $m$-dimensional
orientable distributions $D$ in $\Tan {\cal M}$ and the set of the
equivalence classes $\{ {\cal Y}\}_{\cal M}$
 of non-vanishing, locally decomposable
$m$-multivector fields in ${\cal M}$.

If ${\cal Y}\in\vf^m({\cal M})$ is non-vanishing and locally decomposable, and
$U\subseteq {\cal M}$ is a connected open set, the distribution associated
with the class $\{ {\cal Y}\}_U$ is denoted by ${\cal D}_U({\cal Y})$.
If $U={\cal M}$ we write ${\cal D}({\cal Y})$.

A non-vanishing, locally decomposable
multivector field ${\cal Y}\in\vf^m({\cal M})$ is said to be {\sl integrable}
(resp. {\sl involutive}) if
 its associated distribution ${\cal D}_U({\cal Y})$ is integrable
(resp. involutive).
Of course, if ${\cal Y}\in\vf^m({\cal M})$ is integrable (resp. involutive),
then so is every other in its equivalence class $\{ {\cal Y}\}$,
and all of them have the same integral manifolds.
Moreover, {\sl Frobenius theorem} allows us to say that
a non-vanishing and locally decomposable multivector field is integrable
 if, and only if, it is involutive.
Nevertheless, in many applications we have locally decomposable
multivector fields ${\cal Y}\in\vf^m({\cal M})$ 
which are not integrable in ${\cal M}$,
 but integrable in a submanifold of ${\cal M}$.
A (local) algorithm for finding this submanifold
 has been developed \cite{EMR-98}.

The particular situation in which we are interested
is the study of multivector fields in fiber bundles.
If $\pi\colon {\cal M}\to M$ is a fiber bundle,
we will be interested in the case where the integral manifolds of
integrable multivector fields in ${\cal M}$ are sections of $\pi$.
Thus, ${\cal Y}\in\vf^m({\cal M})$ is said to be {\sl $\pi$-transverse}
if, at every point $y\in {\cal M}$,
$(\inn ({\cal Y})(\pi^*\beta))_y\not= 0$, for every $\beta\in\df^m(M)$
with $\beta (\pi(y))\not= 0$.
Then, if ${\cal Y}\in\vf^m({\cal M})$ is integrable,
it is $\pi$-transverse if, and only if,
its integral manifolds are local sections of $\pi\colon {\cal M}\to M$.
In this case, if $\phi\colon U\subset M\to {\cal M}$
is a local section with $\phi (x)=y$ and $\phi (U)$ is
the integral manifold of ${\cal Y}$ through $y$,
then $\Tan_y({\rm Im}\,\phi)={\cal D}_y({\cal Y})$.

Finally, it is clear that  classes of locally decomposable and
$\pi$-transverse multivector fields $\{{\cal Y}\} \subseteq \vf^m ({\cal M})$
are in one-to-one correspondence with orientable Ehresmann connection
forms $\nabla$ in $\pi\colon{\cal M}\to M$. This correspondence is
characterized by the fact that the horizontal subbundle associated
with $\nabla$ is ${\cal D}({\cal Y})$. In this correspondence,
classes of integrable locally decomposable and $\pi$-transverse
$m$ multivector fields correspond to flat orientable Ehresmann
connections.

\subsection{Hamiltonian multivector fields in multisymplectic manifolds}

(See \cite{Ca-96a} and \cite{LMS-2003} for details).

Let ${\cal M}$ be a $n$-dimensional differentiable manifold and
$\Omega\in\df^{m+1}({\cal M})$.
The couple $({\cal M},\Omega)$ is said to be a
{\sl multisymplectic manifold} if $\Omega$ is closed and $1$-nondegenerate;
that is, for every $p\in{\cal M}$, and $X_p\in\Tan_p{\cal M}$,
 we have that $\inn(X_p)\Omega_p=0$ if, and only if, $X_p=0$.

If $({\cal M},\Omega)$ is a multisymplectic manifold,
${\cal X}\in\vf^k({\cal M})$ is said to be a
{\sl Hamiltonian $k$-multivector field}
if $\inn ({\cal X})\Omega$ is an exact $(m+1-k)$-form; that is,
there exists $\zeta \in\df^{m-k}({\cal M})$ such that
\beq
\inn ({\cal X})\Omega =\d\zeta
\label{ham}
\eeq
$\zeta$ is defined modulo closed $(m-k)$-forms.  The class
$\{\zeta\}\in\df^{m-k}({\cal M})/Z^{m-k}({\cal M})$
defined by $\zeta$ is called the
{\sl Hamiltonian} for ${\cal X}$, and every element in this class
$\hat\zeta\in\{\zeta\}$
is said to be a {\sl Hamiltonian form} for ${\cal X}$.
Furthermore, ${\cal X}$ is said to be a
{\sl locally Hamiltonian $k$-multivector field}
if $\inn ({\cal X})\Omega$ is a closed $(m+1-k)$-form.
In this case, for every point $x\in{\cal M}$, there is an open
neighbourhood $W\subset{\cal M}$ and $\zeta \in\df^{m-k}(W)$ such that
$$
\inn ({\cal X})\Omega = \d\zeta \qquad \mbox{\rm (on $W$)}
$$
As above, changing ${\cal M}$ by $W$, we obtain the
Hamiltonian for ${\cal X}$, $\{\zeta\}\in\df^{k-m-1}(W)/Z^{k-m-1}(W)$,
and the {\sl local Hamiltonian forms} for ${\cal X}$.

Conversely, $\zeta \in\df^k({\cal M})$ (resp. $\zeta \in\df^k(W)$)
is said to be a {\sl Hamiltonian $k$-form} (resp.
a {\sl local Hamiltonian $k$-form}) if there exists a multivector field
${\cal X}\in\vf^{m-k}({\cal M})$ (resp. ${\cal X}\in\vf^{m-k}(W)$)
such that (\ref{ham}) holds (resp. on $W$).
In particular, when $k=0$, that is, if $\zeta \in\Cinfty({\cal M})$),
then the existence of Hamiltonian $m$-multivector fields
for $\zeta$ is assured (see \cite{Ca-96a}).

\subsection{Multimomentum bundles}

(See, for instance, \cite{EMR-00}).

Let $\pi\colon E\to M$ be the {\sl configuration bundle} of
a field theory, (with $\dim\, M=m$, $\dim\, E=n+m$).
There are several multimomentum bundle structures associated with it.

First we have
$\Lambda_2^m\Tan^*E$, which is the bundle of $m$-forms on
$E$ vanishing by the action of two $\pi$-vertical vector fields.
Furthermore, if $J^1\pi\to E\to M$ denotes the {\sl first-order jet bundle}
over $E$, the set made of the affine maps from $J^1\pi$ to
$\Lambda^m\Tan^*M$, denoted as ${\rm Aff}(J^1\pi,\Lambda^m\Tan^*M)$,
is another bundle over $E$ which is canonically diffeomorphic to
$\Lambda_2^m\Tan^*E$ \cite{CCI-91}, \cite{EMR-00}.
We will denote
$$
{\cal M}\pi\equiv\Lambda_2^m\Tan^*E\simeq{\rm Aff}(J^1\pi,\Lambda^m\Tan^*M)
$$
It is called the {\sl extended multimomentum bundle},
and its canonical submersions are denoted
$$
\kappa\colon{\cal M}\pi\to E \quad ; \quad
\bar\kappa=\pi\circ\kappa\colon{\cal M}\pi\to M
$$

${\cal M}\pi$ is a subbundle of $\Lambda^m\Tan^*E$,
the multicotangent bundle of $E$ of order $m$
(the bundle of $m$-forms in $E$).
Then ${\cal M}\pi$ is endowed with canonical forms.
First we have the ``tautological form''
$\Theta\in\df^m({\cal M}\pi)$ which is defined as follows:
let $(x,\alpha )\in\Lambda_2^m\Tan^*E $, with $x\in E$ and
$\alpha\in\Lambda_2^m\Tan_x^*E$; then,
for every $X_1,\ldots,X_m\in\Tan_{(x,\alpha)}({\cal M}\pi)$,
$$
\Theta ((x,\alpha );\moment{X}{1}{m}):=
\alpha (x;\Tan_{(x,\alpha)}\kappa(X_1),\ldots ,\Tan_{(q,\alpha)}\kappa(X_m))
$$
Thus we define the multisymplectic form
 $$
\Omega:=-\d\Theta\in\df^{m+1}({\cal M}\pi)
$$
They are known as the {\sl multimomentum Liouville $m$ and $(m+1)$-forms}.

We can introduce natural coordinates in ${\cal M}\pi$
adapted to the bundle $\pi\colon E\to M$, which
are denoted by $(x^\nu,y^A,p^\nu_A,p)$, and such that
$\omega=\d^mx$.
Then the local expressions of these forms are
\beq
 \Theta=p^\nu_A\d y^A\wedge\d^{m-1}x_\nu+p\d^mx
 \quad , \quad
 \Omega=-\d p^\nu_A\wedge\d y^A\wedge\d^{m-1}x_\nu-\d p\wedge\d^mx
 \label{coor1}
\eeq
(where $d^{m-1}x_\nu:=\inn\left(\derpar{}{x^\nu}\right)\d^mx$).

Consider $\Lambda^m_1\Tan^*E\equiv\pi^*\Lambda^m\Tan^*M$,
which is another bundle over $E$, whose sections are the $\pi$-semibasic
$m$-forms on $E$, and denote by $J^1\pi^*$ the quotient
$\Lambda_2^m\Tan^*E/\Lambda_1^m\Tan^*E\equiv{\cal M}\pi/\Lambda_1^m\Tan^*E$.
We have the natural submersions
 $$
 \tau\colon J^1\pi^*\to E \quad ;\quad
 \bar\tau=\pi\circ\tau\colon J^1\pi^*\to M
 $$
Furthermore, the natural submersion $\mu\colon{\cal M}\pi\to J^1\pi^*$
endows ${\cal M}\pi$ with the structure of an affine bundle over $J^1\pi^*$,
with $\tau^*\Lambda^m_1\Tan^*E$ as the associated vector bundle.
$J^1\pi^*$ is usually called the {\sl restricted multimomentum bundle}
associated with the bundle $\pi\colon E\to M$.

Natural coordinates in $J^1\pi^*$
(adapted to the bundle $\pi\colon E\to M$) are denoted by $(x^\nu,y^A,p^\nu_A)$.

We have the diagram
$$
\begin{array}{ccc}
{\cal M}\pi &
\begin{picture}(135,20)(0,0)
\put(65,8){\mbox{$\mu$}}
\put(0,3){\vector(1,0){135}}
\end{picture}
& J^1\pi^*
\\ &
\begin{picture}(135,100)(0,0)
\put(34,84){\mbox{$\kappa$}}
\put(93,82){\mbox{$\tau$}}
\put(7,55){\mbox{$\bar\kappa$}}
 \put(115,55){\mbox{$\bar\tau$}}
 \put(58,30){\mbox{$\pi$}}
\put(65,55){\mbox{$E$}}
 \put(65,0){\mbox{$M$}}
\put(0,102){\vector(3,-2){55}}
 \put(135,102){\vector(-3,-2){55}}
\put(0,98){\vector(2,-3){55}}
 \put(135,98){\vector(-2,-3){55}}
\put(70,48){\vector(0,-1){35}}
\end{picture} &
\end{array}
$$

Hamiltonian systems can be defined in ${\cal M}\pi$ or in $J^1\pi^*$.
The construction of the Hamiltonian formalism in $J^1\pi^*$ was
pionered in \cite{CCI-91} (see also \cite{EMR-99b} and \cite{EMR-00}), 
while a formulation in ${\cal M}\pi$ has been stated recently
\cite{PR-2002}.
In the following sections we review the main concepts of the formalism 
in $J^1\pi^*$, and we make an extensive development 
of the formalism in ${\cal M}\pi$.

\section{Hamiltonian systems in $J^1\pi^*$}
 \protect\label{hs}

First we consider the standard definition of
Hamiltonian systems in field theory,
which is stated using the restricted multimomentum bundle
$J^1\pi^*$.

\subsection{Restricted Hamiltonian systems}
\protect\label{rhs}

 \begin{definition}
 Consider the bundle $\bar\tau\colon J^1\pi^*\to M$.
\ben
\item
 A section $h\colon J^1\pi^*\to{\cal M}\pi$ of the projection
 $\mu$ is called a {\rm Hamiltonian section} of $\mu$.
\item
 The differentiable forms
 $$
 \Theta_{h}:=h^*\Theta \quad ,\quad
 \Omega_{h}:=-\d\Theta_{h}=h^*\Omega
 $$
 are called the {\rm Hamilton-Cartan $m$ and $(m+1)$ forms} of $J^1\pi^*$
 associated with the Hamiltonian section $h$.
\item
 The couple $\hsjpi$ is said to be a {\rm restricted Hamiltonian system},
(or just a {\rm Hamiltonian system}).
\een
\end{definition}

 In a local chart of natural coordinates,
 a Hamiltonian section is specified by a {\sl local Hamiltonian function}
 ${\rm h}\in\Cinfty (U)$, $U\subset J^1\pi^*$, such that
 $h(x^\nu,y^A,p^\nu_A)\equiv
 (x^\nu,y^A,p^\nu_A,p=-{\rm h}(x^\gamma,y^B,p_B^\eta))$.
 The local expressions of the Hamilton-Cartan forms
 associated with $h$ are
\beq
 \Theta_h = p_A^\nu\d y^A\wedge\d^{m-1}x_\nu -{\rm h}\d^mx
 \quad , \quad
 \Omega_h = -\d p_A^\nu\wedge\d y^A\wedge\d^{m-1}x_\nu +
 \d {\rm h}\wedge\d^mx
\label{omegaH}
 \eeq

\begin{remark} {\rm
Notice that
$\Omega_h$ is $1$-nondegenerate; that is, a multisymplectic
form (as a simple calculation in coordinates shows).
} \end{remark}

Hamiltonian sections can be obtained from connections.
In fact, if we have a connection $\nabla$ in $\pi\colon E\to M$,
it induces a linear section $h^\nabla\colon J^1\pi^*\to{\cal M}\pi$
of $\mu$ \cite{CCI-91}.
Then, if $\Theta$ is the canonical $m$-form in $\df^m({\cal M}\pi)$, the forms
 \beq
 \Theta_{h^\nabla}:= h^{\nabla*}\Theta\in\df^m(J^1\pi^*)
 \quad ,\quad
 \Omega_{h^\nabla}:=-\d\Theta_{h^\nabla}\in\df^{m+1}(J^1\pi^*)
 \label{splitomega}
\eeq
 are the {\sl Hamilton-Cartan $m$ and $(m+1)$ forms} of $J^1\pi^*$
 associated with the connection $\nabla$.
In a system of natural coordinates in $J^1\pi^*$, if
 \(\displaystyle\nabla=\d x^\nu\otimes\left(
 \derpar{}{x^\nu}+{\mit\Gamma}_\nu^A\derpar{}{y^A}\right)\)
is the local expression of the connection $\nabla$,
 the local expressions of these Hamilton-Cartan forms
 associated with $\nabla$ are
 \beann
 \Theta_{h^\nabla} &=& p^\nu_A(\d y^A-{\mit\Gamma}^A_\eta\d x^\eta )
 \wedge\d^{m-1}x_\nu =
 p^\nu_A\d y^A\wedge
 \d^{m-1}x_\nu-p^\nu_A{\mit\Gamma}^A_\nu\d^mx
\nonumber \\
 \Omega_{h^\nabla} &=&
 -\d p^\nu_A\wedge\d y^A\wedge\d^{m-1}x_\nu+
 {\mit\Gamma}^A_\nu\d p^\nu_A\wedge
 \d^mx+ p^\nu_A\d{\mit\Gamma}^A_\nu\wedge\d^mx
 \eeann
Observe that a local Hamiltonian function associated with
$h^\nabla$ is ${\rm h}^\nabla=p^\nu_A{\mit\Gamma}^A_\nu$.

\subsection{Variational principle and field equations}
 \protect\label{vpfe}

Now we establish the field equations for restricted Hamiltonian
systems. They can be derived from a variational principle. In
fact, first we state:

 \begin{definition}
 Let $\hsjpi$ be a restricted Hamiltonian system.
 Let $\Gamma(M,J^1\pi^*)$ be
 the set of sections of $\bar\tau$. Consider the map
 $$
 \begin{array}{ccccc}
 {\bf H}&\colon&\Gamma(M,J^1\pi^*)&\longrightarrow&\Real
 \\
 & &\psi&\mapsto&\int_M\psi^*\Theta_{h}
 \end{array}
 $$
(where the convergence of the integral is assumed).
 The {\rm variational problem} for this restricted Hamiltonian system
 is the search for the critical (or
 stationary) sections of the functional ${\bf H}$,
 with respect to the variations of $\psi$ given
 by $\psi_t =\sigma_t\circ\psi$, where $\{\sigma_t\}$ is the
 local one-parameter group of any compact-supported
 $Z\in\vf^{{\rm V}(\bar\tau)}(J^1\pi^*)$
 (where $\vf^{{\rm V}(\bar\tau)}(J^1\pi^*)$ denotes
 the module of $\bar\tau$-vertical vector fields in $J^1\pi^*$), that is:
 $$
 \frac{d}{d t}\Big\vert_{t=0}\int_M\psi_t^*\Theta_{h} = 0
 $$
 This is the so-called {\rm Hamilton-Jacobi principle}
 of the Hamiltonian formalism.
 \label{hjvp}
 \end{definition}

Then the following fundamental theorem is proven (see also
\cite{EMR-00}):

 \begin{teor}
 Let $\hsjpi$ be a restricted Hamiltonian system.
 The following assertions on a
 section $\psi\in\Gamma(M,J^1\pi^*)$ are equivalent:
 \ben
 \item
 $\psi$ is a critical section for the variational problem posed by
the Hamilton-Jacobi principle.
\item
$\psi^*\inn (Z)\Omega_{h}= 0$, for every 
$Z\in\vf^{{\rm V}(\bar\tau)}(J^1\pi^*)$.
 \item
 $\psi^*\inn (X)\Omega_{h}= 0$, for every
 $X\in\vf (J^1\pi^*)$.
 \item
 If $(U;x^\nu ,y^A,p_A^\nu )$ is a natural system of
 coordinates in $J^1\pi^*$, then
 $\psi$ satisfies the following system of equations in $U$
 \beq
 \derpar{(y^A\circ\psi)}{x^\nu}=
 \derpar{{\rm h}}{p^\nu_A}\circ\psi\equiv
 \derpar{{\rm h}}{p^\nu_A}\Big\vert_\psi
 \quad ;\quad
 \derpar{(p_A^\nu\circ\psi)}{x^\nu}=
 - \derpar{{\rm h}}{y^A}\circ\psi\equiv
 - \derpar{{\rm h}}{y^A}\Big\vert_\psi
 \label{HDWnocov}
 \eeq
where ${\rm h}$ is a local Hamiltonian function associated with $h$.
 They are known as the
 {\rm Hamilton-De Donder-Weyl equations} of the restricted Hamiltonian system.
 \een
 \label{equics}
 \end{teor}
\proof
\qquad ($1\ \Longleftrightarrow 2$)\quad
We assume that $\partial U$ is a $(m-1)$-dimensional manifold
and that $\bar\tau(supp\,(Z))\subset U$,
for every compact-supported $Z\in\vf^{{\rm V}(\bar\tau)}(J^1\pi^*)$.
Then
\beann
\frac{d}{d t}\Big\vert_{t=0}\int_U\psi_t^*\Theta_h  &=&
\frac{d}{d t}\Big\vert_{t=0}\int_U\psi^*(\sigma_t^*\Theta_h) =
\int_U\psi^*\left(\lim_{t\to 0}
\frac{\sigma_t^*\Theta_h-\Theta_h}{t}\right)
\\ &=&
\int_U\psi^*(\Lie(Z)\Theta_h) =
\int_U\psi^*(\inn(Z)\d\Theta_h+\d\inn(Z)\Theta_h)
\\ &=&
-\int_U\psi^*(\inn(Z)\Omega_h-\d\inn(Z)\Theta_h)=
-\int_U\psi^*(\inn(Z)\Omega_h)+
\int_U\d[\psi^*(\inn(Z)\Theta_h)]
\\ &=&
-\int_U\psi^*(\inn(Z)\Omega_h)+
\int_{\partial U}\psi^*(\inn(Z)\Theta_h)=
-\int_U\psi^*(\inn(Z)\Omega_h)
\eeann
(as a consequence of Stoke's theorem and the hypothesis made on the
supports of the vertical fields).
Thus, by the fundamental theorem of the variational calculus
we conclude that 
\(\displaystyle\frac{d}{d t}\Big\vert_{t=0}\int_U\psi_t^*\Theta_h=0\)
$\Leftrightarrow$ $\psi^*(\inn(Z)\Omega_h)=0$,
for every compact-supported $Z\in\vf^{{\rm V}(\bar\tau)}(J^1\pi^*)$.
However, as compact-supported vector fields generate locally the
$\Cinfty(J^1\pi^*)$-module of vector fields in $J^1\pi^*$,
it follows that the last equality holds for every
$Z\in\vf^{{\rm V}(\bar\tau)}(J^1\pi^*)$.

\qquad ($2\ \Longleftrightarrow 3$)\quad
If ${\rm p}\in{\rm Im}\,\psi$, then
$\Tan_{\rm p}J^1\pi^*=
{\rm V}_{\rm p}(\bar\tau)\oplus\Tan_{\rm p}({\rm Im}\,\psi)$.
So if $X\in\vf(J^1\pi^*)$, then
$$
X_{\rm p}=(X_{\rm p}-\Tan_{\rm p}(\psi\circ\bar\tau)(X_{\rm p}))+
\Tan_{\rm p}(\psi\circ\bar\tau)(X_{\rm p})\equiv
X^V_{\rm p}+X^{\psi}_{\rm p}
$$
and therefore
$$
\psi^*(\inn(X)\Omega_h)=
\psi^*(\inn(X^V)\Omega_h)+
\psi^*(\inn(X^{\psi})\Omega_h)=
\psi^*(\inn(X^{\psi})\Omega_h)=0
$$
since $\psi^*(\inn(X^V)\Omega_h)=0$ by the above item,
and furthermore,
$X^{\psi}_{\rm p}\in\Tan_{\rm p}({\rm Im}\,\psi)$,
and $\dim\,({\rm Im}\,\psi)=m$, being
$\Omega_h\in\df^{m+1}(J^1\pi^*)$.
Hence we conclude that
 $\psi^*(\inn (X)\Omega_h)= 0$, for every
 $X\in\vf (J^1\pi^*)$.
The converse is proved reversing this reasoning.

 \quad\quad ($3\ \Leftrightarrow\ 4$)\quad
If \(\displaystyle X=\alpha^\nu\derpar{}{x^\nu}+
 \beta^A\derpar{}{y^A}+\gamma^\nu_A\derpar{}{p^\nu_A}\in\vf(J^1\pi^*)\) ,
 taking into account the local expression (\ref{omegaH})
of $\Omega_h$, we have
 \beann
 \inn (X)\Omega_h &=&
 (-1)^\eta\alpha^\eta\left(\d p^\nu_A\wedge\d y^A\wedge\d^{m-2}x_{\eta\nu}-
 \derpar{{\rm h}}{p^\nu_A} \d p^\nu_A \wedge\d^{m-1}x_\eta\right)
 \\ &+&
 \beta^A\left(\d p^\nu_A\wedge\d^{m-1}x_\nu +
 \derpar{{\rm h}}{y^A}\d^mx \right) +
 \gamma^\nu_A \left(-\d y^A\wedge\d^{m-1}x_\nu +
 \derpar{{\rm h}}{p^\nu_A}\wedge\d^mx\right)
 \eeann
 but if $\psi =(x^\nu ,y^A(x^\eta ),p^\nu_A(x^\eta ))$, then
\beann
 \psi^*\inn (X)\Omega_h  &=&
 (-1)^{\eta+\nu}\alpha^\eta\left(\derpar{(y^A\circ\psi)}{x^\nu}-
 \derpar{{\rm h}}{p^\nu_A}\Big\vert_\psi\right)
\derpar{(p^\nu_A\circ\psi)}{x^\eta} \d^mx
 \\ & &
+ \beta^A\left(\derpar{(p^\nu_A\circ\psi)}{x^\nu}+
 \derpar{{\rm h}}{y^A}\Big\vert_\psi\right)\d^mx +
 \gamma^\nu_A\left(-\derpar{(y^A\circ\psi)}{x^\nu}+
 \derpar{{\rm h}}{p^\nu_A}\Big\vert_\psi \right)\d^mx
 \eeann
 and, as this holds for every $X\in\vf (J^1\pi^*)$,
 we conclude that $\psi^*\inn (X)\Omega_h=0$ if, and only if,
 the Hamilton-De Donder-Weyl equations hold for $\psi$.
 \qed

\begin{remark} {\rm
 It is important to point out that equations (\ref{HDWnocov})
 are not covariant, since
 the Hamiltonian function ${\rm h}$ is defined only locally,
 and hence it is not intrinsically defined.
 In order to write a set of covariant Hamiltonian
 equations we must use a global Hamiltonian function,
 that is, a {\sl Hamiltonian density} 
(see \cite{CCI-91} and \cite{EMR-00} for comments on this subject).

Observe also that the solution to these equations is not unique.
} \end{remark}

\subsection{Hamiltonian equations for multivector fields}

(See \cite{EMR-98} and  \cite{EMR-2002} for more details).

Let $\hsjpi$ be a restricted Hamiltonian system.
 The problem of finding critical sections solutions of the
 Hamilton-Jacobi principle can be formulated equivalently as follows:
 to find a distribution $D$ of $\Tan (J^1\pi^*)$
 satisfying that:
\bit
\item
 $D$ is $m$-dimensional.
\item
 $D$ is $\bar\tau$-transverse.
\item
 $D$ is integrable (that is, {\sl involutive\/}).
\item
 The integral manifolds of $D$ are the critical sections of
 the Hamilton-Jacobi principle.
\eit

However, as explained in Section \ref{mvfc},
these kinds of distributions are associated with classes of integrable
(i.e., non-vanishing, locally decomposable and involutive) $\bar\tau$-transverse
multivector fields in $J^1\pi^*$.
 The local expression in natural coordinates of an element of one
of these classes is
 \beq
 {\cal X}=\bigwedge_{\nu=1}^m
 f\left(\derpar{}{x^\nu}+F_\nu^A\derpar{}{y^A}+
 G^\rho_{A\nu}\derpar{}{p^\rho_A}\right)
 \label{locmvf2}
 \eeq
  where $f\in\Cinfty(J^1\pi^*)$ is a non-vanishing function.

Therefore, the problem posed by the Hamilton-Jacobi principle
can be stated in the following way
\cite{EMR-99b}, \cite{LMS-2003}:

 \begin{teor}
 Let $\hsjpi$ be a restricted Hamiltonian system,
and $\{{\cal X}\}\subset\vf^m(J^1\pi^*)$
a class of integrable and $\bar\tau$-transverse multivector fields.
 Then, the integral manifolds of $\{{\cal X}\}$ are
 critical section for the variational problem posed by
the Hamilton-Jacobi principle if, and only if,
 \beq
 \inn ({\cal X}_h)\Omega_h=0 \quad ,   \quad
 \mbox{\rm for every ${\cal X}_h\in\{ {\cal X}_h\}$}
 \label{hameq1}
\eeq
  \label{hameq}
\end{teor}

\begin{remark} {\rm
The $\bar\tau$-transversality condition for multivector fields
solution to (\ref{hameq1}) can be stated by demanding that
$\inn({\cal X}_h)(\bar\tau^*\omega)\not= 0$. In particular, if we take
$\inn({\cal X}_h)(\bar\tau^*\omega)= 1$ we are choosing a representative
of the class of $\bar\tau$-transverse multivector fields solution
to (\ref{hameq1}). (This is equivalent to putting $f=1$ in the local
expression (\ref{locmvf2})).

Thus, the problem posed in Definition \ref{hjvp} is equivalent to
looking for a multivector field ${\cal X}_h\in\vf^m(J^1\pi^*)$ such that:
 \ben
 \item
 $\inn ({\cal X}_h)\Omega_h=0$.
\item
 $\inn({\cal X}_h)(\bar\tau^*\omega)= 1$.
 \item
 ${\cal X}_h$ is integrable.
 \een
\label{rem3}
} \end{remark}

 From the conditions 1 and 2, using the local expressions (\ref{omegaH}) of
 $\Omega_h$ and (\ref{locmvf2}) for ${\cal X}_h$, we obtain that $f=1$ and
$$
F^A_\nu = \derpar{{\rm h}}{p_A^\nu} \quad ;\quad
G^\nu_{A\nu}= -\derpar{{\rm h}}{y^A}
$$
and, if $\psi(x) =(x^\nu ,y^A(x^\gamma),p^\nu_A(x^\gamma))$
must be an integral section of ${\cal X}_h$, then
 $$
 \derpar{(y^A\circ\psi)}{x^\nu}=F^A_\nu\circ\psi \quad ; \quad
 \derpar{(p^\rho_A\circ\psi)}{x^\nu}=G^\rho_{A\nu}\circ\psi
 $$
Thus the Hamilton-De Donder-Weyl equations (\ref{HDWnocov}) for $\psi$
are recovered from  (\ref{hameq1}).

\begin{remark} {\rm
Classes of locally decomposable and $\bar\tau$-transverse
 multivector fields are in one-to one correspondence
 with connections in the bundle
 $\bar\tau\colon J^1\pi^*\to M$ (see Section \ref{mvfc}).
Then, it can be proven \cite{EMR-99b} that the condition stated in
Theorem \ref{hameq} is equivalent to finding an integrable connection
$\nabla_h$ in $J^1\pi^*\to M$ satisfying the equation $$
\inn(\nabla_h)\Omega_h=(m-1)\Omega_h $$ whose integral sections
are the critical sections of the Hamilton-Jacobi problem. Of
course, $\nabla_h$ is the connection associated to the class
$\{{\cal X}_h\}$ solution to (\ref{hameq1}), and ${\cal X}_h$ is
integrable if, and only if, the curvature of $\nabla_h$ vanishes
everywhere. 
\label{rem2} } 
\end{remark}

The expression of $\nabla_h$ in coordinates is
 $$
\nabla_h=
 \d x^\nu\otimes \left(\derpar{}{x^\nu}+F_\nu^A\derpar{}{y^A}+
 G^\rho_{A\nu}\derpar{}{p^\rho_A}\right)
$$

 \begin{definition}
 ${\cal X}_h\in\vf^m(J^1\pi^*)$ will be called a
 {\rm Hamilton-De Donder-Weyl (HDW) multivector field}
 for the system $\hsjpi$ if it is
 $\bar\tau$-transverse, locally decomposable and verifies the
 equation $\inn ({\cal X}_h)\Omega_h=0$.
Then, the associated connection $\nabla_h$ is called a
{\rm Hamilton-De Donder-Weyl (HDW) connection} for $\hsjpi$.
 \end{definition}

For restricted Hamiltonian systems, the existence of
 Hamilton-De Donder Weyl multivector fields or connections is guaranteed,
 although they are not necessarily integrable
\cite{EMR-98}, \cite{EMR-2002}:

 \begin{teor}
 {\rm (Existence and local multiplicity of HDW-multivector fields):}
 Let $\hsjpi$ be a restricted Hamiltonian system. Then
there exist classes of HDW-multivector fields $\{ {\cal X}_h\}$.
 In a local system the above solutions depend on $n(m^2-1)$
 arbitrary functions.
 \label{holsecreg2}
 \end{teor}

\begin{remark} {\rm
 In order to find a class of integrable
 HDW-multivector fields (if it exists) we must impose that
${\cal X}_h$ verify the
 integrability condition: the curvature of
$\nabla_h$ vanishes everywhere.
 Hence the number of arbitrary functions will
 in general be less than $n(m^2-1)$.
 If this integrable multivector field 
 does not exist, we can eventually select some particular
 HDW-multivector field solution, and apply an integrability
 algorithm in order to find a
 submanifold ${\cal I}\hookrightarrow J^1\pi^*$ (if it exists),
 where this multivector field is integrable
 (and tangent to ${\cal I}$).
\label{remarkN} }
\end{remark}

\section{Hamiltonian systems in ${\cal M}\pi$}
\protect\label{ehs}

Now we introduce Hamiltonian systems in
the extended multimomentum bundle ${\cal M}\pi$,
and we study their relation with those defined in the above section.

\subsection{Extended Hamiltonian systems}

Now we have the multisymplectic manifold
$({\cal M}\pi,\Omega)$, and we are interested
in defining Hamiltonian systems on this manifold which are
suitable for describing Hamiltonian field theories.
Thus we must consider Hamiltonian or locally Hamiltonian
$m$-multivector fields and forms of a particular kind.
In particular, bearing in mind the requirements in Remark
\ref{rem3}, we can state:

\begin{definition}
The triple $({\cal M}\pi,\Omega,\alpha)$ is said to be
an {\rm extended Hamiltonian system} if:
\ben
\item
$\alpha\in Z^1({\cal M}\pi)$ (it is a closed $1$-form).
\item
There exists a locally decomposable multivector field
 ${\cal X}_\alpha\in\vf^m({\cal M}\pi)$ satisfying that
 \beq
 \inn ({\cal X}_\alpha)\Omega =(-1)^{m+1}\alpha
\quad ,\quad
 \inn ({\cal X}_\alpha)(\bar\kappa^*\omega) = 1
\quad  \mbox{\rm ($\bar\kappa$-transversality)}
 \label{hameq2}
\eeq
\een
If $\alpha$ is an exact form, then $({\cal M}\pi,\Omega,\alpha)$
is an {\rm extended global Hamiltonian system}. In this case, there exist
${\rm H}\in\Cinfty({\cal M}\pi)$ such that $\alpha=\d{\rm H}$,
which are called {\rm Hamiltonian functions} of the system.
(For an extended Hamiltonian system, these functions
exist only locally, and they are called
{\rm  local Hamiltonian functions}).
\label{hameqbis}
\end{definition}

The condition that $\alpha$ is closed plays a crucial role
(see Proposition \ref{alphalocal} and Section \ref{gpehs}).
The factor $(-1)^{m+1}$
in the definition will be justified later 
(see Proposition \ref{orthog} and Remark
\ref{importantremark}).

Observe that, if $({\cal M}\pi,\Omega,\alpha)$ is an
extended global Hamiltonian system, giving a Hamiltonian function
${\rm H}$ is equivalent to giving a Hamiltonian density
$\tilde{\cal H}\equiv{\rm H}(\bar\kappa^*\omega)\in\df^m({\cal M}\pi)$.

 In natural coordinates of ${\cal M}\pi$, the
most general expression for a locally decomposable multivector field
${\cal X}_\alpha\in\vf^m({\cal M}\pi$ is
 \beq
 {\cal X}_\alpha=\bigwedge_{\nu=1}^m
\tilde f \left(\derpar{}{x^\nu}+\tilde F_\nu^A\derpar{}{y^A}+
 \tilde G^\rho_{A\nu}\derpar{}{p^\rho_A}+
 \tilde g_\nu\derpar{}{p} \right)
 \label{locmvf3}
 \eeq
where $\tilde f\in\Cinfty({\cal M}\pi)$ is a non-vanishing function
which is equal to $1$ if the equation
$\inn ({\cal X}_\alpha)(\bar\kappa^*\omega) = 1$ holds.

\begin{remark} {\rm
In addition, 
bearing in mind Remark \ref{remarkN},
the integrability of ${\cal X}_\alpha$ must be imposed.
Then all the multivector fields in the integrable class
$\{ {\cal X}_{\alpha}\}$ have the same integral sections.
\label{5}
} \end{remark}

A first important observation is that not every closed
form $\alpha\in\df^m({\cal M}\pi)$ defines an extended
Hamiltonian system. In fact:

\begin{prop}
If $({\cal M}\pi,\Omega,\alpha)$ is an extended Hamiltonian system,
then $\inn(Y)\alpha\not=0$, for every
$\mu$-vertical vector field $Y\in\vf^{{\rm V}(\mu)}({\cal M}\pi)$, $Y\not=0$.
In particular, for every system of natural coordinates
$(x^\nu,y^A,p_A^\nu,p))$ in ${\cal M}\pi$ 
adapted to the bundle $\pi\colon E\to M$ (with $\omega=\d^mx$),
 \beq
 \inn\left(\derpar{}{p}\right)\alpha=1
\label{ortogonal}
 \eeq
\label{orthog}
\end{prop}
\proof
In order to prove this, we use natural coordinates of ${\cal M}\pi$.
The local expression of $\Omega$ is given in (\ref{coor1}),
and a $\mu$-vertical vector field is locally given by
\(\displaystyle Y=f\derpar{}{p}\).
 Then, if ${\cal X}_\alpha\in\vf^m({\cal M}\pi)$
is a multivector field solution to the equations (\ref{hameq2}),
we have
\beann
\inn(Y)\alpha &=&
(-1)^{m+1}\inn(Y)\inn ({\cal X}_\alpha)\Omega=
(-1)^{m+1}(-1)^m\inn ({\cal X}_\alpha)\inn(Y)\Omega
\\ &=&
-\inn ({\cal X}_\alpha)\inn\left(f\derpar{}{p}\right)
[-\d p\wedge\d^mx-\d p^\nu_A\wedge\d y^A\wedge\d^{m-1}x_\nu]
\\ &=&
-\inn ({\cal X}_\alpha)\inn\left(f\derpar{}{p}\right)[-\d p\wedge\d^mx]=
f\inn ({\cal X}_\alpha)\d^mx = f
\eeann
and, as $Y\not= 0\ \Leftrightarrow\ f\not=0$, the first result holds.
In particular, taking $f=1$, the expression (\ref{ortogonal}) is reached.
\qed

As a consequence of this result we have:

\begin{prop}
If $\hsmpi$ is an extended Hamiltonian system,
locally $\alpha=\d p+\beta$, where $\beta$ is a
closed and $\mu$-basic local $1$-form in ${\cal M}\pi$.
\label{alphalocal}
\end{prop}
\proof
As a consequence of (\ref{ortogonal}), $\alpha=\d p+\beta$ locally,
where $\beta$ is a $\mu$-semibasic local $1$-form.
But, as $\alpha$ is closed, so is $\beta$. Hence,
for every $Y\in\vf^{{\rm V}(\mu)}({\cal M}\pi)$, we have that
$\Lie(Y)\beta=\inn(Y)\d\beta+\d\inn(Y)\beta=0$,
and $\beta$ is $\mu$-basic.
\qed

Therefore, by Poincar\'e's lemma, on an open set $U\subset{\cal M}\pi$
$\alpha$ has necessarily the following coordinate expression
\beq
\alpha= \d p+\d\tilde{\rm h}(x^\nu,y^A,p^\nu_A)
\label{alphacoor}
\eeq
where $\tilde{\rm h}=\mu^*{\rm h}$, for some
${\rm h}\in\Cinfty(\mu(U))$.
Then, if ${\rm H}$ is a (local) Hamiltonian function for $\alpha$;
that is, such that $\alpha=\d{\rm H}$ (at least locally),
we have that (see also \cite{PR-2002})
\beq
{\rm H}= p+\tilde{\rm h}(x^\nu,y^A,p^\nu_A)
\label{Hcoor}
\eeq
where $\tilde{\rm h}(x^\nu,y^A,p^\nu_A)$ is determined up to a constant.

 Conversely, every closed form $\alpha\in\df^1({\cal M}\pi)$
 satisfying the above condition defines
 an extended Hamiltonian system since,
 in an analogous way to Theorem \ref{holsecreg2},
 we can prove:

\begin{teor}
 Let $\alpha\in Z^1({\cal M}\pi)$ satisfying the condition stated in
Propositions \ref{orthog} and \ref{alphalocal}.
 Then there exist locally decomposable multivector fields
 ${\cal X}_\alpha\in\vf^m({\cal M}\pi)$ (not necessarily integrable)
 satisfying equations (\ref{hameq2})
 (and hence $\hsmpi$ is an extended Hamiltonian system).
 In a local system the above solutions depend on $n(m^2-1)$
 arbitrary functions.
 \label{holsecreg3}
 \end{teor}
\proof
 We use the local expressions (\ref{coor1}),
 (\ref{alphacoor}) and (\ref{locmvf3}) for
 $\Omega$, $\alpha$ and ${\cal X}_\alpha$ respectively.
 Then $\inn ({\cal X}_\alpha)(\bar\kappa^*\omega) = 1$
 leads to $\tilde f=1$. Furthermore, from
 $\inn ({\cal X}_\alpha)\Omega=(-1)^{m+1}\alpha$
 we obtain that the equality for the coefficients on
 $\d p_A^\nu$ leads to
 \beq
 \tilde F^A_\nu = \derpar{{\rm H}}{p_A^\nu} = \derpar{\tilde{\rm h}}{p_A^\nu}
 \qquad (\mbox{for every $A,\nu$})
 \label{eqsG1}
 \eeq
 For the coefficients on $\d y^A$ we have
 \beq
 \tilde G^\nu_{A\nu}= -\derpar{{\rm H}}{y^A}= -\derpar{\tilde{\rm h}}{y^A}
 \qquad (A=1,\ldots ,n)
 \label{eqsG2}
 \eeq
 and for the coefficients on $\d x^\nu$, using these results, we obtain
 \bea
 \tilde g_\nu &=&
-\derpar{{\rm H}}{x^\nu}+\tilde F^A_\nu\tilde G^\eta_{A\eta}-
\tilde F^A_\eta\tilde G^\eta_{A\nu}=
-\derpar{{\rm H}}{x^\nu}+\derpar{{\rm H}}{p_A^\nu}\tilde G^\eta_{A\eta}-
\derpar{{\rm H}}{p_A^\eta}\tilde G^\eta_{A\nu}
\nonumber \\ &=&
-\derpar{\tilde{\rm h}}{x^\nu}+
\derpar{\tilde{\rm h}}{p_A^\nu}\tilde G^\eta_{A\eta}-
\derpar{\tilde{\rm h}}{p_A^\eta}\tilde G^\eta_{A\nu}
\qquad (A=1,\ldots ,n;\ \eta\not=\nu)
 \label{eqsG3}
 \eea
where the coefficients $G^\rho_{A\nu}$ are related by the equations
(\ref{eqsG2}). Finally, the coefficient on $\d p$ are identical,
 taking into account the above results.

 Thus, equations (\ref{eqsG1}) make a system of $nm$
 linear equations which determines univocally the functions $\tilde F^A_\nu$,
 equations (\ref{eqsG2}) are a compatible system of
 $n$ linear equations on the $nm^2$ functions $\tilde G^\gamma_{A\nu}$,
 and equations (\ref{eqsG3}) make a system of $m$
 linear equations which determines univocally the functions $\tilde g_\nu$.
 In this way, solutions to equations (\ref{hameq2})
 are determined locally from the relations (\ref{eqsG1}) and (\ref{eqsG3}),
 and through the $n$ independent linear equations (\ref{eqsG2}).
 Therefore, there are $n(m^2-1)$ arbitrary functions.
 These results assure the local existence of ${\cal X}_\alpha$.
 The global solutions are
 obtained using a partition of unity subordinated to a
 covering of ${\cal M}\pi$ made of natural charts.
\qed

(A further local analysis of these multivector fields solution
and other additional details can be found in 
\cite{EMR-99b} and \cite{PR-2002}).

\begin{remark} {\rm
With regard to this result, it is important to point out that,
if ${\cal X}_\alpha\in\vf^m({\cal M}\pi)$ is a
solution (not necessarily integrable) to the equations (\ref{hameq2}), 
then every multivector field ${\cal X}'_\alpha\in\{{\cal X}_\alpha\}$;
that is, such that ${\cal X}'_\alpha=\tilde f{\cal X}_\alpha$
(where $\tilde f\in\Cinfty({\cal M}\pi)$ is non-vanishing)
is a solution to the equations
$$
 \inn ({\cal X}'_\alpha)\Omega =\tilde f(-1)^{m+1}\alpha
\quad ,\quad
 \inn ({\cal X}'_\alpha)(\bar\kappa^*\omega) = \tilde f
\quad  \mbox{\rm ($\bar\kappa$-transversality)}
$$
In particular, if we have a $1$-form $\alpha=\d{\rm H}$ (locally), with
\(\displaystyle 0\not=\derpar{{\rm H}}{p}\not=1\), then the
$\bar\kappa$-transversality condition must be stated as
\(\displaystyle\inn ({\cal X}_\alpha)(\bar\kappa^*\omega)=-\derpar{{\rm H}}{p}\),
and the solutions ${\cal X}_\alpha$ 
to the equation $\inn ({\cal X}_\alpha)\Omega =(-1)^{m+1}\alpha$
have the local expression (\ref{locmvf3})
with \(\displaystyle\tilde f=-\derpar{{\rm H}}{p}\), and the other coefficients
being solutions to the system of equations
$$
\tilde f\tilde F^A_\nu = \derpar{{\rm H}}{p_A^\nu} \quad ,\quad
\tilde f\tilde G^\nu_{A\nu}= -\derpar{{\rm H}}{y^A} \quad ,\quad
\tilde f\tilde g_\nu =
-\derpar{{\rm H}}{x^\nu}+\derpar{{\rm H}}{p_A^\nu}\tilde G^\eta_{A\eta}-
\derpar{{\rm H}}{p_A^\eta}\tilde G^\eta_{A\nu}
\quad (\eta\not=\nu)
$$
\label{importantremark}
} \end{remark}

Thus, in an analogous way to restricted Hamiltonian systems in $J^1\pi^*$,
 we define:

 \begin{definition}
 ${\cal X}_\alpha\in\vf^m({\cal M}\pi)$ will be called an
 {\rm extended Hamilton-De Donder-Weyl multivector field}
 for the system $\hsmpi$ if it is
 $\bar\kappa$-transverse, locally decomposable and verifies the equation
 $\inn ({\cal X}_\alpha)\Omega=(-1)^{m+1}\alpha$.
 Then, the associated connection $\nabla_\alpha$
 in the bundle $\bar\kappa\colon{\cal M}\pi\to M$
 is called an {\rm extended Hamilton-De Donder-Weyl connection} for $\hsmpi$.
 \end{definition}

 Now, if $\{ X_\alpha\}$ is integrable and
 $\tilde\psi(x) =(x^\nu,y^A(x^\gamma),p^\nu_A(x^\gamma),p(x^\gamma))$
 must be an integral section of ${\cal X}_\alpha$ then
 \beq
 \derpar{(y^A\circ\tilde\psi)}{x^\nu}=\tilde F^A_\nu\circ\tilde\psi  \quad , \quad
 \derpar{(p^\rho_A\circ\tilde\psi)}{x^\nu}=\tilde G^\rho_{A\nu}\circ\tilde\psi
 \quad , \quad \derpar{(p\circ\tilde\psi)}{x^\nu}=\tilde g_\nu\circ\tilde\psi
 \label{eqsG4}
 \eeq
 so equations (\ref{eqsG1}), (\ref{eqsG2}) and (\ref{eqsG3})
 give PDE's for $\tilde\psi$.
 In particular, the Hamilton-De Donder-Weyl equations (\ref{HDWnocov})
 are recovered from (\ref{eqsG1}) and (\ref{eqsG2}).

As for restricted Hamiltonian systems,
 in order to find a class of integrable extended
 HDW-multivector fields (if it exists) we must impose that
 ${\cal X}_\alpha$ verify the integrability condition, that is,
 that the curvature of $\nabla_\alpha$ vanishes everywhere,
 and thus the number of arbitrary functions will
 in general be less than $n(m^2-1)$.
 Just as in that situation, we cannot assure                     
 the existence of an integrable solution.
 If it does not exist, we can eventually select some particular
 extended HDW-multivector field solution,
 and apply an integrability algorithm in order to find a
 submanifold of ${\cal M}\pi$ (if it exists),
 where this multivector field is integrable (and tangent to it).

\subsection{Geometric properties of extended Hamiltonian systems}
\protect\label{gpehs}

Most of the properties of the extended Hamiltonian systems
are based in the following general results:

\begin{lem}
Let $\mu\colon {\cal M}\to{\cal F}$ be a surjective submersion, with
$\dim\, {\cal M}=\dim\,{\cal F}+r$. Consider
$\moment{\alpha}{1}{r}\in\df^1({\cal M})$ such that
$\alpha\equiv\alpha_1\wedge\ldots\wedge\alpha_r$ is a
closed $r$-form, and $\alpha({\bf p})\not=0$,
for every ${\bf p}\in {\cal M}$. 
Finally, let $\{\alpha\}^0:=\{ Z\in\vf({\cal M})\,\vert\, \inn(Z)\alpha=0\}$
be the annihilator of $\alpha$.
Therefore:
\ben
\item
$\{\alpha\}^0:=\{ Z\in\vf({\cal M})\,\vert\, \inn(Z)\alpha_i=0,\ 
\forall i=1\ldots r\}$.
\item
$\{\alpha\}^0$ generates an involutive distribution in ${\cal M}$
of corank equal to $r$, which is called the
{\rm characteristic distribution} of $\alpha$, and is denoted
${\cal D}_\alpha$.

\noindent
If, in addition, the condition
$\inn(Y)\alpha\not=0$ holds, for every $Y\in\vf^{{\rm V}(\mu)}({\cal M})$, then:
\item
${\cal D}_\alpha$ is a $\mu$-transverse distribution.
\item
The integral submanifolds of ${\cal D}_\alpha$ are
$r$-codimensional and $\mu$-transverse local submanifolds of ${\cal M}$.
\item
$\Tan_{\bf p}{\cal M}={\rm V}_{\bf p}(\mu)\oplus\{\alpha\}_{\bf p}^0$,
for every ${\bf p}\in {\cal M}$.
\item
If $S$ is an integral submanifold of ${\cal D}_\alpha$, then
$\mu\vert_S\colon S\to{\cal F} $ is a local diffeomorphism.
\item
For every integral submanifold $S$ of ${\cal D}_\alpha$,
and ${\bf p}\in S$, there exists $W\subset{\cal M}$, 
with ${\bf p}\in W$, such that $h=(\mu\vert_{W\cap S})^{-1}$
is a local section of $\mu$ defined on
$\mu(W\cap S)$ (which is an open set of ${\cal F}$).
\een
\label{previous}
\end{lem}
\proof
First, observe that, for every ${\bf p}\in {\cal M}$,
$\alpha({\bf p})\not=0$ implies that
$\alpha_i({\bf p})$, for every $i=1,\ldots,r$,
are linearly independent, then
$$
0=\inn(Z)\alpha=
\sum_{i=1}^r(-1)^{i-1}\inn(Z)\alpha_i
(\alpha_1\wedge\ldots\wedge\alpha_{i-1}\wedge\alpha_{i+1}
\wedge\ldots\wedge\alpha_r)
\Leftrightarrow \inn(Z)\alpha_i=0
$$
hence the statement in the item 1 holds and, as a consequence,
we conclude that $\{\alpha\}^0$ generates a distribution in ${\cal M}$
of rank equal to $\dim\,{\cal F}$.

Furthermore, if $\alpha$ is closed,
for every $Z_1,Z_2\in\{\alpha\}^0$, we obtain that
$[Z_1,Z_2]\in\{\alpha\}^0$ because
$$
\inn([Z_1,Z_2])\alpha=\Lie(Z_1)\inn(Z_2)\alpha-\inn(Z_2)\Lie(Z_1)\alpha=
-\inn(Z_2)[\inn(Z_1)\d\alpha-\d\inn(Z_1)\alpha]=0
$$
Then ${\cal D}_\alpha$ is involutive.

The other properties follow straighforwardly from these results and the condition
$\inn(Y)\alpha\not=0$, for every $Y\in\vf^{{\rm V}(\mu)}({\cal M})$.
\qed

Now, from this lemma we have that:

\begin{prop}
If $\hsmpi$ is an extended Hamiltonian system, then:
\ben
\item
${\cal D}_\alpha$ is a $\mu$-transverse involutive distribution
of corank equal to $1$.
\item
The integral submanifolds $S$ of ${\cal D}_\alpha$ are
$1$-codimensional and $\mu$-transverse local submanifolds of ${\cal M}\pi$.
(We denote by $\jmath_S\colon S\hookrightarrow{\cal M}\pi$ the natural embedding).
\item
For every ${\bf p}\in{\cal M}\pi$, we have that
$\Tan_{\bf p}{\cal M}\pi={\rm V}_{\bf p}(\mu)\oplus({\cal D}_\alpha)_{\bf p}$,
and thus, in this way, $\alpha$ defines an integrable connection
in the affine bundle $\mu\colon{\cal M}\pi\to J^1\pi^*$.
\item
If $S$ is an integral submanifold of ${\cal D}_\alpha$, then
$\mu\vert_S\colon S\to J^1\pi^*$ is a local diffeomorphism.
\item
For every integral submanifold $S$ of ${\cal D}_\alpha$,
and ${\bf p}\in S$, there exists $W\subset{\cal M}\pi$, 
with ${\bf p}\in W$, such that $h=(\mu\vert_{W\cap S})^{-1}$
is a local Hamiltonian section of $\mu$ defined on
$\mu(W\cap S)$.
\een
\end{prop}

\begin{remark} {\rm
Observe that, if $\hsmpi$ is an extended Hamiltonian
system, as $\alpha=\d{\rm H}$ (locally), every local Hamiltonian
function ${\rm H}$ is a constraint defining locally the integral
submanifolds of ${\cal D}_\alpha$. Thus, bearing in mind the
coordinate expression (\ref{Hcoor}), the local Hamiltonian sections
associated with these submanifolds are locally expressed as
 $$
h(x^\nu,y^A,p^\nu_A)=
(x^\nu,y^A,p^\nu_A, p=-{\rm h}(x^\gamma,y^B,p_B^\eta))
 $$
where $\mu^*{\rm h}=\tilde{\rm h}$.
} 
\end{remark}

A relevant question is under what conditions the existence of global
Hamiltonian sections is assured.
The answer is given by the following:

\begin{prop}
Let $\hsmpi$ be an extended global Hamiltonian system.
If there is a Hamiltonian function ${\rm H}\in\Cinfty({\cal M}\pi)$,
and $k\in\Real$, such that
$\mu({\rm H}^{-1}(k))=J^1\pi^*$, then
there exists a global Hamiltonian section
$h\in\Gamma(J^1\pi^*,{\cal M}\pi)$.
\end{prop}
\proof
In order to construct $h$, we prove that,
for every ${\rm q}\in J^1\pi^*$, we have that
$\mu^{-1}({\rm q})\cap S_k$ contains only one point.
Let $(U;x^\nu,y^A,p^\nu_A,p)$ be a local chart
in ${\cal M}\pi$, with ${\rm q}\in\mu(U)$.
By Proposition \ref{alphalocal} we have that
${\rm H}\vert_U=p+\mu^*{\rm h}$,
for ${\rm h}\in\Cinfty(\mu(U))$.
If ${\bf p}\in\mu^{-1}({\rm q})\cap S_k$ we have that
$$
k={\rm H}({\bf p})=p({\bf p})-(\mu^*{\rm h})({\bf p}=
p({\bf p})-{\rm h}(\mu({\bf p}))
$$
then $p({\bf p})$  is determined by ${\rm q}$, and 
${\bf p}$ is unique.
This allows to define a global section
$h\colon J^1\pi^*\to{\cal M}\pi$ by
$h({\rm q}):=\mu^{-1}({\rm q})\cap S_k$,
for every ${\rm q}\in J^1\pi^*$,
which obviously does not depend on the local charts considered.
\qed

Observe that, if the first de Rahm cohomology group
$H^1({\cal M}\pi)=0$, then every extended Hamiltonian system is a global one,
but this does not assure the existence of global Hamiltonian sections,
as we have shown.

In addition, we have:

\begin{prop}
Given an extended Hamiltonian system $\hsmpi$,
every extended HDW multivector field ${\cal X}_\alpha\in\vf^m({\cal M}\pi)$
for the system $\hsmpi$ is tangent to every
integral submanifold of ${\cal D}_\alpha$. 
As a consequence, if ${\cal X}_\alpha$ is integrable, 
then its integral sections are contained in the
integral submanifolds of ${\cal D}_\alpha$.
 \label{tangent}
\end{prop}
\proof
 By definition, an extended HDW multivector field is locally decomposable,
 so locally ${\cal X}_\alpha=X_1\wedge\ldots\wedge X_m$,
 with $X_1,\ldots,X_m\in\vf({\cal M}\pi)$.
 Then ${\cal X}_\alpha$ is tangent to every integral submanifold
 $S$ of ${\cal D}_\alpha$ if, and only if, $X_\nu$
 are tangent to $S$, for every $\nu=1,\ldots,m$.
 But, as ${\cal D}_\alpha$ is the characteristic distribution of $\alpha$,
 this is equivalent to $\jmath_S^*\inn(X_\nu)\alpha=0$,
 and this is true because
 $$
\inn(X_\nu)\alpha=\inn(X_\nu)\inn({\cal X}_\alpha)\Omega=
\inn(X_\nu)\inn(X_1\wedge\ldots\wedge X_m)\Omega=0
 $$
The last consequence is immediate.
\qed

\begin{remark} {\rm
Observe that, if ${\cal X}_\alpha=X_1\wedge\ldots\wedge X_m$
locally, using the local expressions (\ref{locmvf3}) and (\ref{alphacoor})
and equations (\ref{eqsG1}) and (\ref{eqsG2}),
the conditions $\inn(X_\nu)\alpha=0$ lead to
\beann
0 &=&
\derpar{\tilde{\rm h}}{x^\nu}+\tilde F^A_\nu\derpar{\tilde{\rm h}}{y^A}+
\tilde G^\rho_{A\nu}\derpar{\tilde{\rm h}}{p_A^\rho}+\tilde g_\nu
\\ &=&
\derpar{\tilde{\rm h}}{x^\nu}-
\derpar{\tilde{\rm h}}{p_A^\rho}\tilde G^\rho_{A\rho}+
\tilde G^\rho_{A\nu}\derpar{\tilde{\rm h}}{p_A^\rho}+\tilde g_\nu
\\ &=&
\derpar{\tilde{\rm h}}{x^\nu}-
\derpar{\tilde{\rm h}}{p_A^\rho}(\tilde G^\eta_{A\eta}+\tilde G^\nu_{A\nu})+
\tilde G^\eta_{A\nu}\derpar{\tilde{\rm h}}{p_A^\eta}+
\tilde G^\nu_{A\nu}\derpar{\tilde{\rm h}}{p_A^\nu}+\tilde g_\nu
\\ &=&
\derpar{\tilde{\rm h}}{x^\nu}-
\derpar{\tilde{\rm h}}{p_A^\nu}\tilde G^\eta_{A\eta}+
\derpar{\tilde{\rm h}}{p_A^\eta}\tilde G^\eta_{A\nu}+\tilde g_\nu
\qquad (\rho=1,\ldots,m;\ \nu\ {\rm fixed},\ \eta\not=\nu)
\eeann
which are the equations (\ref{eqsG3}). So these equations are
consistency conditions.
(See also the comment in Remark \ref{r2}).
}
\label{r1}
\end{remark}

Finally, we have the following result:

\begin{prop}
The integral submanifolds of ${\cal D}_\alpha$
are $m$-coisotropic submanifolds of $({\cal M}\pi,\Omega)$.
\label{coiso}
\end{prop}
\proof
Let $S$ be an integral submanifold of ${\cal D}_\alpha$.
First remember that, for every ${\bf p}\in S$,
the {\sl $m$-orthogonal multisymplectic complement} of $S$ at ${\bf p}$ is
the vector space
$$
\Tan_{\bf p}S^{\perp ,m}:=\{ X_{\bf p}\in\Tan_{\bf p}{\cal M}\pi\, \vert\,
\inn(X_{\bf p}\wedge{\cal X}_{\bf p})\Omega_{\bf p}=0,\
\mbox{\rm for every ${\cal X}_{\bf p}\in\bigwedge^m\Tan_{\bf p}S\equiv
\bigwedge^m({\cal D}_\alpha)_{\bf p}$}\}
$$
and $S$ is said to be a {\sl $m$-coisotropic submanifold}
of $({\cal M}\pi,\Omega)$ if $\Tan_{\bf p}S^{\perp ,m}\subset\Tan_{\bf p}S$
\cite{CIL-98}, \cite{LMS-2003}.
Then, for every $X_{\bf p}\in\Tan_{\bf p}S^{\perp,m}$,
if ${\cal X}_\alpha$ is a HDW-multivector field for the
extended Hamiltonian system $\hsmpi$,
as $({\cal X}_\alpha)_{\bf p}\in\Lambda^m\Tan_{\bf p}S$,
by Proposition \ref{tangent} we have
$$
0=\inn(X_{\bf p})\inn(({\cal X}_\alpha)_{\bf p})\Omega_{\bf p}=
\inn(X_{\bf p})\alpha_{\bf p}
$$
therefore $X_{\bf p}\in({\cal D}_\alpha)_{\bf p}=\Tan_{\bf p}S$.
\qed

This statement generalizes a well-known result in time-dependent
mechanics (see the example in section \ref{nads}): considering the
line bundle
$\mu\colon\Tan^*Q\times\Tan^*\Real\to\Tan^*Q\times\Real$, the zero
section gives a canonical coisotropic embedding of the submanifold
$\Tan^*Q\times\Real$ into the symplectic manifold
$\Tan^*(Q\times\Real)\simeq\Tan^*Q\times\Tan^*\Real$. Furthermore,
in field theories, every maximal integral submanifold $S$ of
${\cal D}_\alpha$ gives a local $m$-coisotropic embedding of
$U\subset\mu(S)\subset J^1\pi^*$ into ${\cal M}\pi$, given by
$(\mu\vert_S)^{-1}$, which is obviously not canonical.

\subsection{Relation between extended and restricted Hamiltonian systems}
\protect\label{rerhs}

Now we can establish the relation between extended and restricted
Hamiltonian systems in $J^1\pi^*$. Taking into account the
considerations made in the above section, we can state:

\begin{teor}
Let $\hsmpi$ be an extended global Hamiltonian system, and $\hsjpi$ a
restricted Hamiltonian system such that ${\rm Im}\,h=S$
is an integral submanifold of ${\cal D}_\alpha$.
 For every ${\cal X}_\alpha\in\vf^m({\cal M}\pi)$ 
solution to the equations (\ref{hameq2}):
$$
\inn ({\cal X}_\alpha)\Omega=(-1)^{m+1}\alpha \quad , \quad
\inn ({\cal X}_\alpha)(\bar\kappa^*\omega)=1
$$
there exists ${\cal X}_h\in\vf^m(J^1\pi^*)$ which
is $h$-related with ${\cal X}_\alpha$
and is a solution to the equations
$$
\inn ({\cal X}_h)\Omega_h=0 \quad , \quad
\inn ({\cal X}_h)(\bar\tau^*\omega)=1
$$
(i.e., satisfying the conditions $1$ and $2$ in Remark \ref{rem3}).

 Furthermore, if ${\cal X}_\alpha$ is integrable, then
${\cal X}_h$ is integrable too, and the integral sections of ${\cal X}_h$
are recovered from those of ${\cal X}_\alpha$ as follows:
if $\tilde\psi\colon M\to{\cal M}\pi$ is an integral
section of ${\cal X}_\alpha$, then 
$\psi=\mu\circ\tilde\psi\colon M\to J^1\pi^*$
is an integral section of ${\cal X}_h$.
\label{related}
\end{teor}
\proof
 Given $S={\rm Im}\,h$,
 let $\jmath_S \colon S\hookrightarrow{\cal M}\pi$ be the natural
 embedding, and $h_S\colon J^1\pi^*\to S$ the
 diffeomorphism between $J^1\pi^*$ and ${\rm Im}\,h$,
 then $h=\jmath_S\circ h_S$.

 If ${\cal X}_\alpha\in\vf^m({\cal M}\pi)$ is a solution
 to the equations (\ref{hameq2}), by Proposition \ref{tangent},
 it is tangent to $S$, then there exists ${\cal X}_S\in\vf^m(S)$
 such that $\Lambda^m\jmath_{S*}{\cal X}_S={\cal X}_\alpha\vert_S$.
 Let ${\cal X}_h\in\vf^m(J^1\pi^*)$ defined by
 ${\cal X}_h=\Lambda^mh_{S*}^{-1}{\cal X}_S$. Therefore, from the equation
 $\inn ({\cal X}_\alpha)\Omega=(-1)^{m+1}\alpha$
and the condition $\jmath_S^*\alpha=0$
(which holds because $S$ is an integral submanifold of ${\cal D}_\alpha$), 
we obtain
 \beann
 0 &=& h^*[\inn ({\cal X}_\alpha)\Omega-(-1)^{m+1}\alpha]=
 (\jmath_S\circ h_S)^*[\inn ({\cal X}_\alpha)\Omega-(-1)^{m+1}\alpha]=
 (h_S^*\circ\jmath_S^*) [\inn ({\cal X}_\alpha)\Omega-(-1)^{m+1}\alpha]
\\ &=&
 h_S^*[\inn ({\cal X}_S)\jmath_S^*\Omega-(-1)^{m+1}\jmath_S^*\alpha]=
 \inn ({\cal X}_h)(h_S^*\circ\jmath_S^*)\Omega=
 \inn ({\cal X}_h)h^*\Omega=\inn ({\cal X}_h)\Omega_h
 \eeann
Furthermore, bearing in mind that $\mu\circ h={\rm Id}_{J^1\pi^*}$,
we have that
 $$
 \inn ({\cal X}_h)(\bar\tau^*\omega)=
 \inn ({\cal X}_h)[(\bar\kappa\circ h)^*\omega]=
 h^*[\inn ({\cal X}_\alpha)(\bar\kappa^*\omega)]
 $$
and, if $\inn ({\cal X}_\alpha)(\bar\kappa^*\omega)=1$,
this equality holds, in particular, at the points of
the image of $h$, therefore
$\inn ({\cal X}_h)(\bar\tau^*\omega)=1$.
Then ${\cal X}_h$ is the desired multivector field,
since ${\cal X}_h\vert_S={\cal X}_S=\Lambda^m_*{\cal X}_h$.

Finally, if ${\cal X}_\alpha$ is integrable,
as it is tangent to $S$, the integral sections 
of ${\cal X}_\alpha$ passing
through any point of $S$ remain in $S$, and hence
they are the integral sections of ${\cal X}_S$,
so ${\cal X}_\alpha$ is integrable and,
as a consequence, ${\cal X}_h$ is integrable too.
 \qed

All of these properties lead to establish the following:

\begin{definition}
Given an extended global Hamiltonian system $\hsmpi$,
and considering all the Hamiltonian sections
 $h\colon J^1\pi^*\to{\cal M}\pi$ such that
 ${\rm Im}\,h$ are integral submanifolds of ${\cal D}_\alpha$,
 we have a family $\{ (J^1\pi^*,h)\}_\alpha$, which will be called the
 {\rm class of restricted Hamiltonian systems associated with $\hsmpi$}.
\end{definition}

As it is obvious, in general, the above result holds only locally.

The following result show how to obtain extended Hamiltonian systems from
restricted Hamiltonian ones, at least locally. In fact:

\begin{prop}
Given a restricted Hamiltonian system $\hsjpi$,
let $\jmath_S\colon S={\rm Im}\, h\hookrightarrow{\cal M}\pi$
be the natural embedding. Then, there exists a unique local form
$\alpha\in\df^1({\cal M}\pi)$ such that:
\ben
\item
$\alpha\in\ Z^1({\cal M}\pi)$ (it is a closed form).
\item
$\jmath_S^*\alpha=0$.
\item
$\inn(Y)\alpha\not=0$, for every non-vanishing
$Y\in\vf^{{\rm V}(\mu)}({\cal M}\pi)$ and, in particular, such that
 \(\displaystyle\inn\left(\derpar{}{p}\right)\alpha=1\),
 for every system of natural coordinates $(x^\nu,y^A,p_A^\nu,p)$
 in ${\cal M}\pi$, adapted to the bundle 
$\pi\colon E\to M$ (with $\omega=\d^mx$).
 \een
\label{existuniq}
\end{prop}
\proof
Suppose that there exist $\alpha,\alpha'$ satisfying the above conditions.
Taking into account the comments after Proposition \ref{orthog},
we have that, locally in $U\subset{\cal M}\pi$,
$\alpha=\d p+\beta$ and $\alpha'=\d p+\beta'$,
where $\beta=\mu^*\bar\beta$, $\beta'=\mu^*\bar\beta'$,
with $\bar\beta,\bar\beta'\in B^1(\mu(U))$ (they are exact $1$-forms).
From condition 2 in the statement we have that
$\jmath_S^*\alpha=\jmath_S^*\alpha'$; hence
\beann
0=\jmath_S^*(\alpha-\alpha') &\Longleftrightarrow&
0=\jmath_S^*(\beta-\beta')=(\mu\circ\jmath_S)^*(\bar\beta-\bar\beta')\\
&\Longrightarrow&
0=(\mu\circ\jmath_S\circ h_S)^*(\bar\beta-\bar\beta')\ \Longrightarrow\
\bar\beta-\bar\beta'=0 \ \Longleftrightarrow\ \bar\beta=\bar\beta'
\ \Longleftrightarrow\ \bar\alpha=\bar\alpha'
\eeann
since $\mu\circ\jmath_S\circ h_S=\mu\circ h={\rm Id}_{\mu(U)}$.
This proves the uniqueness.

The existence is trivial since, locally, every section $h$ of $\mu$
is given by a function ${\rm h}\in\Cinfty(\mu(U))$ such that
$p=-{\rm h}(x^\nu,y^A,p_A^\nu)$. Hence
$\alpha\vert_{\mu(U)}=\d p+\d(\mu^*{\rm h})\equiv\d p+\d\tilde{\rm h}$.
\qed

\begin{definition}
Given a restricted Hamiltonian system $\hsjpi$, let
$\alpha\in\df^1({\cal M}\pi)$ be the local form satisfying the
conditions in the above proposition. 
The couple $({\cal M}\pi,\alpha)$ will be called the
(local) {\rm extended Hamiltonian system associated with $\hsjpi$}.
\end{definition}

As a consequence of the last proposition, 
if $\alpha=\d p+\mu^*\bar\beta$, there exists a class
$\{{\rm h}\}\in\Cinfty(\mu(U))/\Real$, such that
$\bar\beta=\d{\rm h}$, where ${\rm h}$ is a representative
of this class. Then:

\begin{corol}
Let $\alpha$ be the unique local $1$-form verifying the conditions
of Proposition \ref{existuniq}, associated with a section $h$.
Consider its characteristic distribution ${\cal D}_\alpha$,
and let $\{h\}_\alpha$ the family of local sections of $\mu$
such that ${\rm Im}\,h\equiv S$ are local integral
submanifolds of ${\cal D}_\alpha$.
Then, for every $h'\in\{ h\}_\alpha$, we have that ${\rm Im}\,h'$ is locally a
level set of the function ${\rm H}=p+\mu^*{\rm h}\equiv p+\tilde{\rm h}$.
\end{corol}
\proof
If $S={\rm Im}\,h'$ then, for every ${\rm p}\in S$,
we have $\Tan_{\rm p}S=({\cal D}_\alpha)_{\rm p}$,
which is equivalent to $\d (p+\mu^*{\rm h})\vert_{\Tan_{\rm p}S}=0$,
and this holds if, and only if, 
${\rm H}\vert_S\equiv(p+\mu^*{\rm h})\vert_S=ctn.$
\qed

Bearing in mind these considerations, we can finally prove that:

\begin{prop}
Let $\hsmpi$ be an extended Hamiltonian system, and
$\{ (J^1\pi^*,h)\}_\alpha$ the class of restricted Hamiltonian systems 
associated with $\hsmpi$. Consider the submanifolds $\{ S_h={\rm Im}\, h\}$,
for every Hamiltonian section $h$ in this class,
and let $\jmath_{S_h}\colon S_h\hookrightarrow{\cal M}\pi$ be
the natural embeddings.
Then the submanifolds $(S_h,\jmath_{S_h}^*\Omega)$
are multisymplectomorphic.
\end{prop}
\proof
Let $h_1,h_2\in\{ h\}$ and $S_1={\rm Im}\, h_1$, $S_2={\rm Im}\, h_2$.
We have the diagram
$$
\begin{array}{ccccc}
&
\begin{picture}(5,60)(0,0)
 \put(0,0){\mbox{$S_1$}}
 \put(-6,52){\mbox{${\cal M}\pi$}}
 \put(2,11){\vector(0,1){35}}
 \put(5,25){\mbox{$\jmath_{S_1}$}}
\end{picture}
&
\begin{picture}(110,60)(0,0)
 \put(52,-38){\vector(-2,3){55}}
 \put(63,-38){\vector(2,3){55}}
\put(25,10){\mbox{$h_1$}}
\put(81,10){\mbox{$h_2$}}
\end{picture}
&
\begin{picture}(10,52)(0,0)
 \put(3,0){\mbox{$S_2$}}
 \put(0,52){\mbox{${\cal M}\pi$}}
 \put(10,11){\vector(0,1){35}}
 \put(-5,25){\mbox{$\jmath_{S_2}$}}
\end{picture}
&
\\
&
\begin{picture}(10,40)(0,0)
\put(0,35){\vector(-3,-2){53}}
\put(-21,12){\mbox{$\mu_1$}}
\end{picture}
&
\begin{picture}(110,40)(0,0)
 \put(6,12){\mbox{$h_{S_1}$}}
\put(47,0){\vector(-3,2){53}}
 \put(68,0){\vector(3,2){53}}
  \put(105,12){\mbox{$h_{S_2}$}}
\end{picture}
&
\begin{picture}(10,40)(0,0)
\put(15,35){\vector(3,-2){53}}
\put(24,12){\mbox{$\mu_2$}}
\end{picture}
&
\\
\begin{picture}(65,10)(0,0)
\put(0,0){\mbox{$J^1\pi^*$}}
\put(25,3){\vector(1,0){110}}
\put(75,6){\mbox{${\rm Id}$}}
\put(67,104){\vector(-2,-3){58}}
\put(30,70){\mbox{$\mu$}}
\end{picture}
& &
\begin{picture}(110,10)(0,0)
\put(45,0){\mbox{$J^1\pi^*$}}
\end{picture}
& &
\begin{picture}(65,10)(0,0)
\put(50,0){\mbox{$J^1\pi^*$}}
\put(45,3){\vector(-1,0){110}}
\put(-15,6){\mbox{${\rm Id}$}}
\put(0,104){\vector(2,-3){58}}
\put(30,70){\mbox{$\mu$}}
\end{picture}
 \end{array}
 $$

Denote $\Omega_1=\jmath_{S_1}^*\Omega$, $\Omega_2=\jmath_{S_2}^*\Omega$.
As a consequence of the above corollary, if
$\Omega_{h_1}=h_1^*\Omega$, $\Omega_{h_2}=h_2^*\Omega$,
we have that $\Omega_{h_1}=\Omega_{h_2}$. But
$$
\Omega_{h_1}=\Omega_{h_2} \Longleftrightarrow
(\jmath_{S_1}\circ h_{S_1})^*\Omega=(\jmath_{S_2}\circ h_{S_2})^*\Omega 
\Longleftrightarrow h_{S_1}^*\Omega_1=h_{S_2}^*\Omega_2 
$$
Then, the map $\Phi:=h_{S_2}\circ\mu_1\colon S_1\to S_2$
is a multisymplectomorphism. In fact, it is obviously a diffeomorphism, and
$$
\Phi^*\Omega_2=(h_{S_2}\circ\mu_1)^*\Omega_2=\mu_1^*h_{S_2}^*\Omega_2=
\mu_1^*h_{S_1}^*\Omega_1=(h_{S_1}\circ\mu_1)^*\Omega_1=\Omega_1
$$
\qed

As an immediate consequence of this,
if ${\cal X}_\alpha\in\vf^m({\cal M}\pi)$ is a solution
to the equations (\ref{hameq2}), the multivector fields
${\cal X}_{S_h}\in\vf^m(S_h)$ such that 
$\Lambda^m(\jmath_{S_h})_*{\cal X}_{S_h}={\cal X}_\alpha\vert_S$,
for every submanifold $S_h$ of this family
(see the proof of Theorem \ref{related}),
are related by these multisymplectomorphisms.

\subsection{Variational principle and field equations}
\protect\label{evpfe}

As in the case of restricted Hamiltonian systems,
the field equations for extended Hamiltonian systems
can be derived from a suitable variational principle.

First, denote by $\vf_\alpha({\cal M}\pi)$ 
the set of vector fields $Z\in\vf({\cal M}\pi)$
which are sections of the subbundle ${\cal D}_\alpha$
of $\Tan{\cal M}\pi$, that is, satisfying that $\inn(Z)\alpha=0$
(and hence, they are tangent to all the integral submanifolds of 
${\cal D}_\alpha$).
Let $\vf_\alpha^{{\rm V}(\bar\kappa)}({\cal M}\pi)\subset
\vf_\alpha({\cal M}\pi)$
be those which are also $\bar\kappa$-vertical.

Furthermore, as we have seen in previous sections,
the image of the sections $\tilde\psi\colon M\to{\cal M}\pi$,
which are solutions to the extended field equations,
must be in the integral submanifolds of the characteristic
distribution ${\cal D}_\alpha$; that is, they are also
integral submanifolds, and hence $\jmath_{\tilde\psi}^*\alpha=0$
(where $\jmath_{\tilde\psi}\colon{\rm Im\,\tilde\psi}\hookrightarrow{\cal M}\pi$
is the natural embedding).
We will denote by $\Gamma_\alpha(M,{\cal M}\pi)$
 the set of sections of $\bar\kappa$ satisfying that
 $\jmath_{\tilde\psi}^*\alpha=0$.

Taking all of this into account, we can state the following:

 \begin{definition}
 Let $\hsmpi$ be an extended Hamiltonian system.
 Consider the map
 $$
 \begin{array}{ccccc}
 \tilde{\bf H}_\alpha&\colon&\Gamma_\alpha(M,{\cal M}\pi)&\longrightarrow&\Real
 \\
 & &\tilde\psi&\mapsto&\int_U\tilde\psi^*\Theta
 \end{array}
 $$
(where the convergence of the integral is assumed).
 The {\rm variational problem} for this extended Hamiltonian system
 is the search for the critical (or
 stationary) sections of the functional $\tilde{\bf H}_\alpha$,
 with respect to the variations of $\tilde\psi\in\Gamma_\alpha(M,{\cal M}\pi)$ 
 given by $\tilde\psi_t =\sigma_t\circ\tilde\psi$, where $\{\sigma_t\}$ is the
 local one-parameter group of any compact-supported vector field
 $Z\in\vf_\alpha^{{\rm V}(\bar\kappa)}({\cal M}\pi)$, that is
 $$
 \frac{d}{d t}\Big\vert_{t=0}\int_U\tilde\psi_t^*\Theta = 0
 $$
 This is the {\rm extended Hamilton-Jacobi principle}.
 \label{ehjvp}
 \end{definition}

Observe that, as $\alpha$ is closed, the variation of
the set $\Gamma_\alpha(M,{\cal M}\pi)$ is stable under the action
of $\vf_\alpha^{{\rm V}(\bar\kappa)}({\cal M}\pi)$. In fact;
being $\alpha$ closed, for every
$Z\in\vf_\alpha^{{\rm V}(\bar\kappa)}({\cal M}\pi)$,
we have that $\Lie(Z)\alpha=\inn(Z)\d\alpha+\d\inn(Z)\alpha=0$,
that is, $\sigma_t^*\alpha=\alpha$.
Hence, if $\tilde\psi\in\Gamma_\alpha(M,{\cal M}\pi)$, 
we obtain
$$
\tilde\psi_t^*\alpha=(\sigma_t\circ\tilde\psi)^*\alpha=
\tilde\psi^*\sigma_t^*\alpha=\tilde\psi^*\alpha=0
$$

Then we have the following fundamental theorems:

 \begin{teor}
 Let $\hsmpi$ be an extended Hamiltonian system.
 The following assertions on a
 section $\tilde\psi\in\Gamma_\alpha(M,{\cal M}\pi)$ are equivalent:
 \ben
 \item
 $\tilde\psi$ is a critical section for the variational problem posed by
the extended Hamilton-Jacobi principle.
\item
$\tilde\psi^*\inn (Z)\Omega= 0$, for every
$Z\in\vf_\alpha^{{\rm V}(\bar\kappa)}({\cal M}\pi)$.
 \item
 $\tilde\psi^*\inn (X)\Omega= 0$, for every
 $X\in\vf_\alpha({\cal M}\pi)$.
\item
If $(U;x^\nu,y^A,p_A^\nu,p)$ is a natural system of
 coordinates in ${\cal M}\pi$, then
 $\tilde\psi$ satisfies the following system of equations in $U$
 \beq
 \derpar{(y^A\circ\tilde\psi)}{x^\nu}=
\derpar{\tilde{\rm h}}{p^\nu_A}\circ\tilde\psi \quad ,\quad
 \derpar{(p_A^\nu\circ\tilde\psi)}{x^\nu}=
-\derpar{\tilde{\rm h}}{y^A}\circ\tilde\psi \quad ,\quad
 \derpar{(p\circ\tilde\psi)}{x^\nu}=
-\derpar{(\tilde{\rm h}\circ\tilde\psi)}{x^\nu}
\label{ehdwe}
  \eeq
where $\tilde{\rm h}=\mu^*{\rm h}$, for some
${\rm h}\in\Cinfty(\mu(U))$,  is any function such that
$\alpha\vert_U=\d p+\d\tilde{\rm h}(x^\nu,y^A,p^\nu_A)$.
These are the
{\rm extended Hamilton-De Donder-Weyl equations} of the extended
Hamiltonian system.
\een
 \label{exequics}
 \end{teor}
\proof
\qquad ($1\ \Longleftrightarrow 2$)\quad
We assume that $\partial U$ is a $(m-1)$-dimensional manifold
and that $\bar\kappa(supp\,(Z))\subset U$,
for every compact-supported
$Z\in\vf_\alpha^{{\rm V}(\bar\kappa)}({\cal M}\pi)$. Then
\beann
\frac{d}{d t}\Big\vert_{t=0}\int_U\tilde\psi_t^*\Theta  &=&
\frac{d}{d t}\Big\vert_{t=0}\int_U\tilde\psi^*(\sigma_t^*\Theta) =
\int_U\tilde\psi^*\left(\lim_{t\to 0}
\frac{\sigma_t^*\Theta-\Theta}{t}\right)
\\ &=&
\int_U\tilde\psi^*(\Lie(Z)\Theta) =
\int_U\tilde\psi^*(\inn(Z)\d\Theta+\d\inn(Z)\Theta)
\\ &=&
-\int_U\tilde\psi^*(\inn(Z)\Omega-\d\inn(Z)\Theta)=
-\int_U\tilde\psi^*(\inn(Z)\Omega)+
\int_U\d[\tilde\psi^*(\inn(Z)\Theta)]
\\ &=&
-\int_U\tilde\psi^*(\inn(Z)\Omega)+
\int_{\partial U}\tilde\psi^*(\inn(Z)\Theta)=
-\int_U\tilde\psi^*(\inn(Z)\Omega)
\eeann
(as a consequence of Stoke's theorem and the hypothesis made on the
supports of the vertical fields).
Thus, by the fundamental theorem of the variational calculus
 we conclude that 
\(\displaystyle\frac{d}{d t}\Big\vert_{t=0}\int_U\psi_t^*\Theta=0\)
$\Leftrightarrow$ $\tilde\psi^*(\inn(Z)\Omega)=0$,
for every compact-supported 
$Z\in\vf_\alpha^{{\rm V}(\bar\kappa)}({\cal M}\pi)$.
But, as compact-supported vector fields generate locally the
$\Cinfty({\cal M}\pi)$-module of vector fields in ${\cal M}\pi$,
it follows that the last equality holds for every
$Z\in\vf_\alpha^{{\rm V}(\bar\kappa)}({\cal M}\pi)$.

\qquad ($2\ \Longleftrightarrow 3$)\quad
If ${\bf p}\in{\rm Im}\,\tilde\psi$, 
and $S$ is the integral submanifold of ${\cal D}_\alpha$ passing
through ${\bf p}$, then 
$$
({\cal D}_\alpha)_{\bf p}=
[{\rm V}_{\bf p}(\bar\kappa)\cap({\cal D}_\alpha)_{\bf p}]\oplus
\Tan_{\bf p}({\rm Im}\,\tilde\psi)
$$
So, for every $X\in\vf_\alpha({\cal M}\pi)$,
$$
X_{\bf p}=(X_{\bf p}-\Tan_{\bf p}(\tilde\psi\circ\bar\kappa)(X_{\bf p}))+
\Tan_{\bf p}(\tilde\psi\circ\bar\kappa)(X_{\bf p})\equiv
X^V_{\bf p}+X^{\tilde\psi}_{\bf p}
$$
and therefore
$$
\tilde\psi^*(\inn(X)\Omega)=
\tilde\psi^*(\inn(X^V)\Omega)+
\tilde\psi^*(\inn(X^{\tilde\psi})\Omega)=
\tilde\psi^*(\inn(X^{\tilde\psi})\Omega)=0
$$
since $\tilde\psi^*(\inn(X^V)\Omega)=0$ by the above item,
and furthermore,
$X^{\tilde\psi}_{\bf p}\in\Tan_{\bf p}({\rm Im}\,\tilde\psi)$,
and $\dim\,({\rm Im}\,\tilde\psi)=m$, being
$\Omega\in\df^{m+1}({\cal M}\pi)$.
Hence we conclude that
 $\tilde\psi^*(\inn (X)\Omega)= 0$, for every
 $X\in\vf_\alpha({\cal M}\pi)$.
The converse is proved reversing this reasoning.

\qquad ($3\ \Longleftrightarrow 4$)\quad
The local expression of any $X\in\vf_\alpha({\cal M}\pi)$ is
$$
X=\lambda^\nu\derpar{}{x^\nu}+\beta^A\derpar{}{y^A}+
\gamma^\nu_A\derpar{}{p^\nu_A}-
\left(\alpha^\eta\derpar{\tilde{\rm h}}{x^\eta}+
\beta^B\derpar{\tilde{\rm h}}{y^B}+
\gamma^\eta_B\derpar{\tilde{\rm h}}{p^\eta_B}\right)\derpar{}{p}
$$
then, taking into account the local expression (\ref{coor1})
of $\Omega$, if $\tilde\psi=(x^\nu,y^A(x^\eta),p^\nu_A(x^\eta),p(x^\eta))$,
we obtain
\beann
 \tilde\psi^*\inn (X)\Omega &=&
 \lambda^\eta\left( \derpar{(p\circ\tilde\psi)}{x^\nu}
+\derpar{\tilde{\rm h}}{x^\nu}\Big\vert_{\tilde\psi}
-\sum_{\eta\not=\nu}\left(\derpar{\tilde{\rm h}}{p_A^\nu}\Big\vert_{\tilde\psi}
\derpar{(p^\eta_A\circ\tilde\psi)}{x^\eta}-
\derpar{\tilde{\rm h}}{p_A^\eta}\Big\vert_{\tilde\psi}
\derpar{(p^\eta_A\circ\tilde\psi)}{x^\nu}\right)\right)\d^mx
 \\ & &
+ \beta^A\left(\derpar{(p^\nu_A\circ\tilde\psi)}{x^\nu}+
 \derpar{\tilde{\rm h}}{y^A}\Big\vert_{\tilde\psi}\right)\d^mx +
 \gamma^\nu_A\left(-\derpar{(y^A\circ\tilde\psi)}{x^\nu}+
 \derpar{\tilde{\rm h}}{p^\nu_A}\Big\vert_{\tilde\psi} \right)\d^mx
 \eeann
and, as this holds for every $X\in\vf_\alpha({\cal M}\pi)$
(i.e., for every $\lambda^\eta,\beta^A,\gamma^\nu_A$),
we conclude that $\psi^*\inn(X)\Omega_h=~0$ if, and only if,
\bea
\derpar{(y^A\circ\tilde\psi)}{x^\nu}&=&
\derpar{\tilde{\rm h}}{p^\nu_A}\Big\vert_{\tilde\psi} \quad ,\quad
 \derpar{(p_A^\nu\circ\tilde\psi)}{x^\nu}\ =\ 
-\derpar{\tilde{\rm h}}{y^A}\Big\vert_{\tilde\psi}
\label{number1}
\\  
 \derpar{(p\circ\tilde\psi)}{x^\nu}&=&
-\derpar{\tilde{\rm h}}{x^\nu}\Big\vert_{\tilde\psi}
+\sum_{\eta\not=\nu}\left(\derpar{\tilde{\rm h}}{p_A^\nu}\Big\vert_{\tilde\psi}
\derpar{(p^\eta_A\circ\tilde\psi)}{x^\eta}-
\derpar{\tilde{\rm h}}{p_A^\eta}\Big\vert_{\tilde\psi}
\derpar{(p^\eta_A\circ\tilde\psi)}{x^\nu}\right)
\label{number2}
\eea
and using equations (\ref{number1}) in (\ref{number2}) we obtain
\beann
\derpar{(p\circ\tilde\psi)}{x^\nu}&=&
-\derpar{\tilde{\rm h}}{x^\nu}\Big\vert_{\tilde\psi}
+\derpar{\tilde{\rm h}}{p_A^\nu}\Big\vert_{\tilde\psi}
\left(\derpar{(p^\delta_A\circ\tilde\psi)}{x^\delta}-
\derpar{(p^\nu_A\circ\tilde\psi)}{x^\nu}\right)
-\left(\derpar{\tilde{\rm h}}{p_A^\delta}\Big\vert_{\tilde\psi}
\derpar{(p^\delta_A\circ\tilde\psi)}{x^\nu}-
\derpar{\tilde{\rm h}}{p_A^\nu}\Big\vert_{\tilde\psi}
\derpar{(p^\nu_A\circ\tilde\psi)}{x^\nu}\right)
\\ &=&
-\derpar{\tilde{\rm h}}{x^\nu}\Big\vert_{\tilde\psi}
-\derpar{\tilde{\rm h}}{y^A}\Big\vert_{\tilde\psi}
\derpar{(y^A\circ\tilde\psi)}{x^\nu}
- \derpar{\tilde{\rm h}}{p_A^\delta}\Big\vert_{\tilde\psi}
 \derpar{(p^\delta_A\circ\tilde\psi)}{x^\nu}
\\ &=&
-\derpar{(\tilde{\rm h}\circ\tilde\psi)}{x^\nu}
\qquad (\delta=1,\ldots,m\ ; \ \nu\ {\rm fixed})
\eeann
\qed

\begin{remark} {\rm
It is important to point out that the last group of equations
(\ref{ehdwe}) are consistency conditions with respect
to the hypothesis made on the sections $\tilde\psi$.
In fact, this group of equations leads to
$p\circ\tilde\psi=-\tilde{\rm h}\circ\tilde\psi+ctn.$
or, what means the same thing, 
$\tilde\psi\in\Gamma_\alpha(M,{\cal M}\pi)$.
(See also the comment in Remark \ref{r1}).
The rest of equations (\ref{ehdwe}) are just
the Hamilton-De Donder-Weyl equations (\ref{HDWnocov}) of the restricted case,
since the local expressions of the functions
$\tilde{\rm h}$ and ${\rm h}$ are the same.
}
\label{r2}
\end{remark}

\begin{teor}
 Let $\hsmpi$ be an extended Hamiltonian system,
and ${\cal X}\in\vf^m({\cal M}\pi)$ an integrable multivector field
verifying the condition $\inn({\cal X})(\bar\kappa^*\omega)=1$.
 Then, the integral manifolds of ${\cal X}$ are
 critical section for the variational problem posed by
the extended Hamilton-Jacobi principle if, and only if,
${\cal X}$ satisfies the condition $\inn({\cal X})\Omega=(-1)^{m+1}\alpha$.
\label{teorHDW}
\end{teor}
\proof
\qquad ($\Longleftarrow$)\quad
Let $S$ be an integral submanifold of ${\cal X}$.
By the $\bar\kappa$-transversality condition 
$\inn({\cal X})(\bar\kappa^*\omega)=1$,
$S$ is locally a section of $\bar\kappa$.
Then, for every ${\bf p}\in S$, there are an open set
$U\subset M$, with $\bar\kappa({\bf p})\in U$,
and a local section $\tilde\psi\colon U\subset M\to{\cal M}\pi$
of $\bar\kappa$, such that ${\rm Im}\,\tilde\psi=S\vert_{\bar\kappa^{-1}(U)}$.
Now, let $q\in U$, and $u_1,\ldots ,u_m\in\Tan_qM$,
with $\inn(u_1\wedge\ldots\wedge u_m)(\omega(\bar\kappa({\bf p})))=1$.
Then, there exists $\lambda\in\Real$ such that
$$
\tilde\psi_*(u_1\wedge\ldots\wedge u_m)=\lambda{\cal X}_{(\tilde\psi(q))}
$$
therefore
$$
\inn(\tilde\psi_*(u_1\wedge\ldots\wedge u_m))(\Omega(\tilde\psi(q)))=
\lambda(-1)^{m+1}\alpha(\tilde\psi(q))
$$
Thus, for every $X\in\vf_\alpha({\cal M}\pi)$ we obtain that
$$
\inn(X_{(\tilde\psi(q))})\inn(\tilde\psi_*(u_1\wedge\ldots\wedge u_m)
(\Omega(\tilde\psi(q)))=0
$$
hence $\tilde\psi^*\inn(X)\Omega= 0$, for every
$X\in\vf_\alpha({\cal M}\pi)$, and $\tilde\psi$ is a critical section
by the third item of the last Theorem.

\qquad ($\Longrightarrow$)\quad
Let ${\bf p}\in{\cal M}\pi$,
by the hypothesis there exists a section 
$\tilde\psi\colon M\to{\cal M}\pi$ such that
\ben
\item
$\tilde\psi({\bf p})=\bar\kappa({\bf p})$.
\item
$\tilde\psi$ is a critical section for the 
extended Hamilton-Jacobi variational problem,
that is, $\tilde\psi^*\inn(X)\Omega= 0$, for every
$X\in\vf_\alpha({\cal M}\pi)$.
\item
${\rm Im}\,\tilde\psi$ is an integral submanifold of ${\cal X}$.
\een
Now, let $u_1,\ldots ,u_m\in\Tan_{\bar\kappa({\bf p})}M$,
with $\inn(u_1\wedge\ldots\wedge u_m)\omega(\bar\kappa({\bf p})))=1$.
Then, there exists $\lambda\in\Real$ such that
$$
\tilde\psi_*(u_1\wedge\ldots\wedge u_m)=\lambda{\cal X}_{\bf p}
$$
but the condition imposed to $u_1,\ldots ,u_m$ leads to
$\lambda=1$. Therefore
$$
\inn(\tilde\psi_*(u_1\wedge\ldots\wedge u_m))(\Omega({\bf p}))=
\inn({\cal X}_{\bf p})(\Omega({\bf p}))
$$
Thus, for every $X\in\vf_\alpha({\cal M}\pi)$,
as $\tilde\psi$ is a critical section, we obtain that
$$
\inn(X_{\bf p})\inn({\cal X}_{\bf p})(\Omega({\bf p}))=0
$$
and hence $\inn(X)\inn({\cal X})\Omega=0$,
for every $X\in\vf_\alpha({\cal M}\pi)$.
This implies that
$\inn({\cal X})\Omega=f\alpha$, for some non-vanishing
$f\in\Cinfty({\cal M}\pi)$.
Nevertheless, as $\hsmpi$ is an extended Hamiltonian system,
in any local chart we have that
$\alpha=\d p+\d\tilde{\rm h}(x^\nu,y^A,p^\nu_A)$,
which by the condition $\inn({\cal X})(\bar\kappa^*\omega)=1$,
and bearing in mind the local expression of $\Omega$,
leads to $f=(-1)^{m+1}$.
So the result holds.
\qed

Observe that the extended Hamilton-De Donder-Weyl equations (\ref{ehdwe})
can also be obtained as a consequence of this last theorem, 
taking into account equations (\ref{eqsG1}), (\ref{eqsG2}), (\ref{eqsG3}) 
and (\ref{eqsG4}).

\section{Almost-regular Hamiltonian systems}
\protect\label{arhs}

There are many interesting cases in Hamiltonian field theories
 where the Hamiltonian field equations are established
 not in $J^1\pi^*$, but rather in a submanifold of $J^1\pi^*$
(for instance, when considering the Hamiltonian formalism
associated with a singular Lagrangian).
Next we consider this kind of systems in $J^1\pi^*$,
as well as in ${\cal M}\pi$.

\subsection{Restricted almost-regular Hamiltonian systems}

 \begin{definition}
A {\rm restricted almost-regular Hamiltonian system} is a triple
$\hsjpio$, where:
\ben
\item
${\cal P}$ is a submanifold of $J^1\pi^*$ with $\dim\,{\cal P}>n+m$, and such that,
if $\jmath_{\cal P}\colon{\cal P}\hookrightarrow J^1\pi^*$
denotes the natural embedding, the map
$\tau_{\cal P}=\tau\circ\jmath_{\cal P}\colon{\cal P}\to E$
is a surjective submersion (and hence, so is the map
$\bar\tau_{\cal P}=\bar\tau\circ\jmath_{\cal P}=\pi\circ\tau_{\cal P}
\colon{\cal P}\to M$).
\item
 $h_{\cal P}\colon{\cal P}\to{\cal M}\pi$ satisfies that
$\mu\circ h_{\cal P}=\jmath_{\cal P}$, and it is called a
{\rm Hamiltonian section} of $\mu$ on ${\cal P}$.
\een
 Then, the differentiable forms
 $$
 \Theta_{h_{\cal P}}:=h_{\cal P}^*\Theta \quad ,\quad
 \Omega_{h_{\cal P}}:=-\d\Theta_{h_{\cal P}}=h_{\cal P}^*\Omega
 $$
 are the {\rm Hamilton-Cartan $m$ and $(m+1)$ forms} on ${\cal P}$
 associated with the Hamiltonian section $h_{\cal P}$.
\label{arhsdef}
\end{definition}

We have the diagram
$$
\begin{array}{ccc}
&
\begin{picture}(135,35)(0,0)
\put(0,-15){\vector(3,2){55}}
\put(60,25){\mbox{${\cal M}\pi$}}
\put(85,20){\vector(3,-2){55}}
\put(14,10){\mbox{$h_{\cal P}$}}
\put(110,10){\mbox{$\mu$}}
\end{picture}
&
\\
{\cal P} &
\begin{picture}(135,20)(0,0)
\put(65,8){\mbox{$\jmath_{\cal P}$}}
\put(0,3){\vector(1,0){135}}
\end{picture}
& J^1\pi^*
\\ &
\begin{picture}(135,100)(0,0)
\put(34,84){\mbox{$\tau_{\cal P}$}}
\put(93,82){\mbox{$\tau$}}
\put(7,55){\mbox{$\bar\tau_{\cal P}$}}
 \put(115,55){\mbox{$\bar\tau$}}
 \put(58,30){\mbox{$\pi$}}
\put(65,55){\mbox{$E$}}
 \put(65,0){\mbox{$M$}}
\put(0,102){\vector(3,-2){55}}
 \put(135,102){\vector(-3,-2){55}}
\put(0,98){\vector(2,-3){55}}
 \put(135,98){\vector(-2,-3){55}}
\put(70,48){\vector(0,-1){35}}
\end{picture} &
\end{array}
$$

\begin{remark} {\rm
Notice that $\Omega_{h_{\cal P}}$ is, in general, a $1$-degenerate form
and hence it is premultisymplectic.
This is the main difference with the regular case.
} \end{remark}

Furthermore, if we make the additional assumption that
${\cal P}\to E$ is a fiber bundle,
the Hamilton-Jacobi variational principle of Definition \ref{hjvp}
can be stated in the same way, now
 using sections of $\bar\tau_{\cal P}\colon{\cal P}\to M$, and the form
 $\Theta_{h_{\cal P}}$.
 So we look for sections $\psi_{\cal P}\in\Gamma(M,{\cal P})$ which are
 stationary with respect to the variations given by
 $\psi_t=\sigma_t\circ\psi_{\cal P}$, where $\{\sigma_t\}$ is a local
 one-parameter group of any compact-supported
$\bar\tau_{\cal P}$-vertical vector field
 $Z_{\cal P}\in\vf({\cal P})$; i.e., such that
 $$
 \frac{d}{d t}\Big\vert_{t=0}
\int_M\psi_t^*\Theta_{h_{\cal P}}= 0
 $$
 Then these critical sections will be characterized by the condition
 (analogous to Theorem \ref{equics})
 $$
 \psi_{\cal P}^*\inn (X_{\cal P})\Omega_{h_{\cal P}} = 0
 \quad ; \quad
 \mbox{\rm for every $X_{\cal P}\in\vf ({\cal P})$}
 $$
And, as in the case of restricted Hamiltonian systems
(Theorem \ref{hameq}), we have that:

 \begin{teor}
 The critical sections of the Hamilton-Jacobi principle are
 the integral sections $\psi_{\cal P}\in\Gamma(M,{\cal P})$
 of a class of integrable and
 $\bar\tau_{\cal P}$-transverse multivector fields
 $\{ {\cal X}_{h_{\cal P}}\}\subset\vf^m({\cal P})$ satisfying that
 $$
 \inn ({\cal X}_{h_{\cal P}})\Omega_{h_{\cal P}}=0 \quad ,   \quad
 \mbox{\rm for every ${\cal X}_{h_{\cal P}}\in\{ {\cal X}_{h_{\cal P}}\}$}
 $$
or equivalently, the integral sections of an integrable multivector field
 ${\cal X}_{h_{\cal P}}\in\vf^m({\cal P})$ such that:
 \ben
 \item
 $\inn ({\cal X}_{h_{\cal P}})\Omega_{h_{\cal P}}=0$.
\item
 $\inn({\cal X}_{h_{\cal P}})(\bar\tau_{\cal P}^*\omega)= 1$.
 \een
  \label{hameq0}
\end{teor}

A multivector field
 ${\cal X}_{h_{\cal P}}\in\vf^m({\cal P})$ will be called a
 {\sl Hamilton-De Donder-Weyl multivector field}
 for the system $\hsjpio$ if it is
 $\bar\tau_{\cal P}$-transverse, locally decomposable and verifies the
 equation $\inn ({\cal X}_{h_{\cal P}})\Omega_{h_{\cal P}}=0$.
Then, the associated connection $\nabla_{h_{\cal P}}$,
which is a connection along the submanifold ${\cal P}$
(see \cite{LMM-95}, \cite{LMM-96} and \cite{LMS-2004}), is called a
{\sl Hamilton-De Donder-Weyl connection} for $\hsjpio$,
and satisfies the equation
$$
\inn(\nabla_{h_{\cal P}})\Omega_{h_{\cal P}}=(m-1)\Omega_{h_{\cal P}}
$$

\begin{remark} {\rm
 It should be noted that, as $\Omega_{h_{\cal P}}$ can be $1$-degenerate,
 the existence of the corresponding Hamilton-De Donder-Weyl multivector fields
 for $\hsjpio$ is in general not assured
 except perhaps on some submanifold
 $S$ of ${\cal P}$, where the solution is not unique.
 A geometric algorithm for determining this submanifold $S$
 has been developed \cite{LMMMR-04}.
} \end{remark}

\subsection{Extended almost-regular Hamiltonian systems}

 \begin{definition}
An {\rm extended almost-regular Hamiltonian system} is a triple
$\hsmpio$, such that:
\ben
\item
$\tilde{\cal P}$ is a submanifold of ${\cal M}\pi$ and, if
$\jmath_{\tilde{\cal P}}\colon\tilde{\cal P}\hookrightarrow{\cal M}\pi$
denotes the natural embedding, then:
\ben
\item
$\kappa_{\tilde{\cal P}}=\kappa\circ\jmath_{\tilde{\cal P}}
\colon\tilde{\cal P}\to E$
is a surjective submersion (and hence, so is the map
$\bar\kappa_{\tilde{\cal P}}=\bar\kappa\circ\jmath_{\tilde{\cal P}}=
\pi\circ\kappa_{\tilde{\cal P}}\colon\tilde{\cal P}\to M$).
\item
$(\mu\circ\jmath_{\tilde{\cal P}})(\tilde{\cal P})\equiv{\cal P}$
 is a submanifold of $J^1\pi^*$.
\item
$\tilde{\cal P}={\cal M}\pi\vert_{\mu(\tilde{\cal P})}$;
that is, for every ${\bf p}\in\tilde{\cal P}$ we have that
$\mu^{-1}(\mu({\bf p}))={\bf p}+\Lambda^m_1\Tan_{\kappa({\bf p})}^*E
\subset\tilde{\cal P}$.
\een
\item
$\alpha_{\tilde{\cal P}}\in Z^1(\tilde{\cal P})$
(it is a closed $1$-form in $\tilde{\cal P}$).
\item
There exists a locally decomposable multivector field
 ${\cal X}_{\alpha_{\tilde{\cal P}}}\in\vf^m(\tilde{\cal P})$ satisfying that
 \beq
 \inn ({\cal X}_{\alpha_{\tilde{\cal P}}})\Omega_{\tilde{\cal P}}
 =(-1)^{m+1}\alpha_{\tilde{\cal P}}
\quad ,\quad
\inn ({\cal X}_{\alpha_{\tilde{\cal P}}})(\bar\kappa_{\tilde{\cal P}}^*\omega)=1
\quad  \mbox{\rm ($\bar\kappa_{\tilde{\cal P}}$-transversality)}
 \label{hameq20}
\eeq
where $\Omega_{\tilde{\cal P}}=\jmath_{\tilde{\cal P}}^*\Omega$.
\een
If $\alpha_{\tilde{\cal P}}$ is an exact form, then $\hsmpio$
is an {\rm extended almost-regular global Hamiltonian system}.
In this case there exist functions
${\rm H}_{\tilde{\cal P}}\in\Cinfty(\tilde{\cal P})$,
which are called {\rm Hamiltonian functions} of the system,
such that
 $\alpha_{\tilde{\cal P}}=\d{\rm H}_{\tilde{\cal P}}$.
(For an extended Hamiltonian system, these functions
exist only locally, and they are called
{\rm  local Hamiltonian functions}).
\label{arehsdef}
\end{definition}

\begin{remark} {\rm
As straighforward consequences of this definition we have that:
\bit
\item
The condition (1.c) of Definition \ref{arehsdef}
imply, in particular, that $\dim\,\tilde{\cal P}>\dim\,E+1$.
Furthermore, it
means that $\tilde{\cal P}$ is the union of fibers of $\mu$.
\item
$\kappa\circ\jmath_{\tilde{\cal P}}$ is a surjective submersion
if, and only if, so is $\tau\circ\mu\circ\jmath_{\tilde{\cal P}}$.
This means that ${\cal P}\equiv{\rm Im}\,(\mu\circ\jmath_{\tilde{\cal P}})$
is a submanifold verifying the conditions stated in the first item of
definition \ref{arhsdef}, and such that
$\dim\,{\cal P}=\dim\,\tilde{\cal P}-1$,
as a consequence of the properties given in item 1
of Definition \ref{arehsdef}.
This submanifold is diffeomorphic to
$\tilde{\cal P}/\Lambda^m_1\Tan^*E$.
\eit
\label{remint}
} \end{remark}

Denoting $\mu_{\tilde{\cal P}}=\mu\circ\jmath_{\tilde{\cal P}}
\colon\tilde{\cal P}\to J^1\pi^*$, and
$\tilde\mu_{\tilde{\cal P}}\colon\tilde{\cal P}\to{\cal P}$
its restriction to the image
(that is, such that 
$\mu_{\tilde{\cal P}}=\jmath_{\cal P}\circ\tilde\mu_{\tilde{\cal P}}$), 
we have the diagram
$$
\begin{array}{ccc}
{\cal P} & 
\begin{picture}(135,25)(0,0)
\put(65,12){\mbox{$\jmath_{\cal P}$}}
\put(0,7){\vector(1,0){135}}
\end{picture}
& J^1\pi^*
\\
\begin{picture}(10,20)(0,0)
\put(5,-10){\vector(0,1){30}}
\put(8,5){\mbox{$\tilde\mu_{\tilde{\cal P}}$}}
\end{picture} &
 \begin{picture}(135,20)(0,0)
\put(55,13){\mbox{$\mu_{\tilde{\cal P}}$}}
\put(0,-17){\vector(3,1){135}}
\end{picture} &
\begin{picture}(10,20)(0,0)
\put(5,-10){\vector(0,1){30}}
\put(8,5){\mbox{$\mu$}}
\end{picture}
\\
\tilde{\cal P} &
\begin{picture}(135,20)(0,0)
\put(65,8){\mbox{$\jmath_{\tilde{\cal P}}$}}
\put(0,3){\vector(1,0){135}}
\put(135,50){\vector(-2,-3){55}}
\put(108,23){\mbox{$\tau$}}
\end{picture}
& {\cal M}\pi
\\ &
\begin{picture}(135,100)(0,0)
\put(34,84){\mbox{$\kappa_{\tilde{\cal P}}$}}
\put(103,87){\mbox{$\kappa$}}
\put(7,55){\mbox{$\bar\kappa_{\tilde{\cal P}}$}}
 \put(115,55){\mbox{$\bar\kappa$}}
 \put(58,30){\mbox{$\pi$}}
\put(65,55){\mbox{$E$}}
 \put(65,0){\mbox{$M$}}
\put(0,102){\vector(3,-2){55}}
 \put(135,102){\vector(-3,-2){55}}
\put(0,98){\vector(2,-3){55}}
 \put(135,98){\vector(-2,-3){55}}
\put(70,48){\vector(0,-1){35}}
\end{picture} &
\end{array}
$$

\begin{remark} {\rm
In addition, as for extended Hamiltonian systems
(see Remarks \ref{remarkN} and \ref{5}),
the integrability of ${\cal X}_{\alpha_{\tilde{\cal P}}}$
is not assured, so it must be imposed. 
Then all the multivector fields in the integrable class
$\{ {\cal X}_{\alpha_{\tilde{\cal P}}}\}$
have the same integral sections.
} \end{remark}

As in Propositions \ref{orthog} and \ref{alphalocal}, we have that:

\begin{prop}
If $\hsmpio$ is an extended almost-regular Hamiltonian system,
then $\inn(Y_{\tilde{\cal P}})\alpha_{\tilde{\cal P}}\not=0$, 
for every non-vanishing $\mu_{\tilde{\cal P}}$-vertical vector field
$Y_{\tilde{\cal P}}\in\vf^{{\rm V}(\mu_{\tilde{\cal P}})}(\tilde{\cal P})$.
In particular, for every system of natural coordinates in
$\tilde{\cal P}$ adapted to the bundle $\pi\colon E\to M$
 (with $\omega=\d^mx$),
 $$
 \inn\left(\derpar{}{p}\right)\alpha_{\tilde{\cal P}}=1
 $$
\label{orthog0}
\end{prop}
\proof
As a consequence of the condition (1.c) of Definition
\ref{arehsdef}, we have the local expression
$$
\Omega_{\tilde{\cal P}}=\jmath_{\tilde{\cal P}}^*\Omega=
\jmath_{\tilde{\cal P}}^*(-\d p^\nu_A\wedge\d y^A\wedge\d^{m-1}x_\nu-
\d p\wedge\d^mx)=
\jmath_{\tilde{\cal P}}^*(-\d p^\nu_A)\wedge\d y^A\wedge\d^{m-1}x_\nu-
\d p\wedge\d^mx
$$
and \(\displaystyle Y_{\tilde{\cal P}}=f\derpar{}{p}\),
for every $\mu$-vertical vector field
in $\tilde{\cal P}$. Therefore, the proof follows
the same pattern as in the proof of
Proposition \ref{orthog}.

The last part of the proof 
is a consequence of the condition (1.c) given in Definition
\ref{arehsdef}, from which we have that every system of natural coordinates in
$\tilde{\cal P}$ adapted to the bundle $\pi\colon E\to M$ contains
the coordinate $p$ of the fibers of $\mu$, and the coordinates 
$(x^\nu)$ in $E$.
This happens because $\tilde{\cal P}$ reduces only degrees of freedom
in the coordinates $p^\nu_A$ of ${\cal M}\pi$.
\qed

\begin{prop}
If $\hsmpio$ is an extended Hamiltonian system,
locally $\alpha_{\tilde{\cal P}}=\d p+\beta_{\tilde{\cal P}}$, 
where $\beta_{\tilde{\cal P}}$ is a
closed and $\tilde\mu_{\tilde{\cal P}}$-basic
 local $1$-form in $\tilde{\cal P}$.
\label{alphalocal0}
\end{prop}

A multivector field
 ${\cal X}_{\alpha_{\tilde{\cal P}}}\in\vf^m(\tilde{\cal P})$ will be called an
 {\sl extended Hamilton-De Donder-Weyl multivector field}
 for the system $\hsmpio$ if it is
 $\bar\kappa_{\tilde{\cal P}}$-transverse, 
locally decomposable and verifies the equation 
$\inn ({\cal X}_{\alpha_{\tilde{\cal P}}})\Omega_{\tilde{\cal P}}=
(-1)^{m+1}\alpha_{\tilde{\cal P}}$.
Then, the associated connection $\nabla_{\alpha_{\tilde{\cal P}}}$ is called a
{\sl Hamilton-De Donder-Weyl connection} for $\hsmpio$,

\begin{remark} {\rm
Notice that $\Omega_{\tilde{\cal P}}$ is usually a $1$-degenerate form
and hence premultisymplectic.

 As a consequence, the existence of extended
 Hamilton-De Donder-Weyl multivector fields
 for $\hsmpio$ is not assured, except perhaps on some submanifold
 $\tilde S$ of ${\cal P}$, where the solution is not unique.
} \end{remark}

\subsection{Geometric properties of extended almost-regular Hamiltonian systems.
Variational principle}

Let $\hsmpio$ be an extended almost-regular Hamiltonian system,
and the submanifold
$\mu_{\tilde{\cal P}}(\tilde{\cal P})\equiv{\cal P}$.
 As for the general case, we can define the 
{\sl characteristic distribution} ${\cal D}_{\alpha_{\tilde{\cal P}}}$
of $\alpha_{\tilde{\cal P}}$. Then, following the same pattern as
in the proofs of the propositions and theorems given in Section
\ref{gpehs} we can prove that:

\begin{prop}
\ben
\item
${\cal D}_{\alpha_{\tilde{\cal P}}}$ is an involutive and
$\mu_{\tilde{\cal P}}$-transverse distribution
of corank equal to $1$.
\item
The integral submanifolds of ${\cal D}_{\alpha_{\tilde{\cal P}}}$ are
 $\mu_{\tilde{\cal P}}$-transverse submanifolds of ${\cal M}\pi$,
with dimension equal to $\dim\,\tilde{\cal P}-1$.
(We denote by $\tilde\jmath_S\colon S\hookrightarrow\tilde{\cal P}$ 
the natural embedding).
\item
For every ${\bf p}\in\tilde{\cal P}$, we have that
$\Tan_{\bf p}\tilde{\cal P}=
{\rm V}_{\bf p}(\mu_{\tilde{\cal P}})\oplus
({\cal D}_{\alpha_{\tilde{\cal P}}})_{\bf p}$,
and thus, in this way, $\alpha_{\tilde{\cal P}}$ defines a connection
in the bundle 
$\mu_{\tilde{\cal P}}\colon\tilde{\cal P}\to{\cal P}$.
\item
If $S$ is an integral submanifold of ${\cal D}_{\alpha_{\tilde{\cal P}}}$, then
$\tilde\mu_{\tilde{\cal P}}\vert_S\colon S\to{\cal P}$
is a local diffeomorphism.
\item
For every integral submanifold $S$ of ${\cal D}_{\alpha_{\tilde{\cal P}}}$,
and ${\bf p}\in S$, there exists $W\subset\tilde{\cal P}$, 
with ${\bf p}\in W$, such that $h=(\mu\vert_{W\cap S})^{-1}$
is a local Hamiltonian section of $\tilde\mu_{\tilde{\cal P}}$ defined on
$\tilde\mu_{\tilde{\cal P}}(W\cap S)$.
\een
\end{prop}

If $\hsmpio$ is an extended almost-regular Hamiltonian system,
as $\alpha_{\tilde{\cal P}}=\d{\rm H}_{\tilde{\cal P}}$ (locally),
every local Hamiltonian function ${\rm H}_{\tilde{\cal P}}$
is a constraint defining
the local integral submanifolds of ${\cal D}_{\tilde{\cal P}}$.
If $\hsmpio$ is an extended almost-regular global Hamiltonian system,
the Hamiltonian functions ${\rm H}_{\tilde{\cal P}}$
are globally defined, and we have:

\begin{prop}
Let $\hsmpio$ be an extended almost-regular global Hamiltonian system.
If there is a global Hamiltonian function 
${\rm H}_{\tilde{\cal P}}\in\Cinfty(\tilde{\cal P})$,
and $k\in\Real$, such that $\mu({\rm H_{\tilde{\cal P}}}^{-1}(k))={\cal P}$,
 then there exists a global Hamiltonian section
$h_{\tilde{\cal P}}\in\Gamma({\cal P},\tilde{\cal P})$.
\end{prop}

\begin{prop}
Given an extended almost-regular Hamiltonian system $\hsmpio$ ,
every extended HDW multivector field 
${\cal X}_{\alpha_{\tilde{\cal P}}}\in\vf^m(\tilde{\cal P})$
for the system $\hsmpio$ is tangent to every
integral submanifold of ${\cal D}_{\alpha_{\tilde{\cal P}}}$.
\end{prop}

At this point, the extended Hamilton-Jacobi variational principle 
of Definition \ref{ehjvp} can be stated in the same way, now
using sections $\psi_{\tilde{\cal P}}$ of 
$\bar\kappa_{\tilde{\cal P}}\colon\tilde{\cal P}\to M$,
satisfying that $\jmath_{\psi_{\tilde{\cal P}}}^*\alpha_{\tilde{\cal P}}=0$
(where $\jmath_{\psi_{\tilde{\cal P}}}\colon
{\rm Im}\,\psi_{\tilde{\cal P}}\hookrightarrow\tilde{\cal P}$
denotes the natural embedding).
Thus, using the notation introduced in section \ref{evpfe}, we look for sections
$\psi_{\tilde{\cal P}}\in\Gamma_{\alpha_{\tilde{\cal P}}}(M,\tilde{\cal P})$
which are stationary with respect to the variations given
 by $\tilde\psi_t =\sigma_t\circ\psi_{\tilde{\cal P}}$,
 where $\{\sigma_t\}$ is a
 local one-parameter group of every compact-supported
$Z\in\vf_{\alpha_{\tilde{\cal P}}}^{{\rm V}(\bar\kappa_{\tilde{\cal P}})}
(\tilde{\cal P})$;
that is
 $$
\frac{d}{d t}\Big\vert_{t=0}
\int_U\tilde\psi_t^*\Theta_{\tilde{\cal P}} = 0
 $$
And then the statements analogous to Theorems \ref{exequics} 
and \ref{teorHDW} can be
established and proven in the present case.

\subsection{Relation between extended and 
restricted almost-regular Hamiltonian systems}
\protect\label{rearrhs}

Finally, we study the relation between extended and 
restricted almost-regular Hamiltonian systems. 
(The proofs of the following propositions and theorems 
are analogous to those in Section \ref{rerhs}).

First, bearing in mind Remark \ref{remint}, we have:

\begin{teor}
Let $\hsmpio$ be an extended global Hamiltonian system, and $\hsjpio$ a
restricted Hamiltonian system such that
$\dim\,\tilde{\cal P}=\dim\,{\cal P}+1$, and ${\rm Im}\,h_{\cal P}=S$
is an integral submanifold of ${\cal D}_{\alpha_{\tilde{\cal P}}}$.
Then, for every ${\cal X}_{\alpha_{\tilde{\cal P}}}\in\vf^m(\tilde{\cal P})$
solution to the equations:
$$
\inn ({\cal X}_{\alpha_{\tilde{\cal P}}})\Omega_{\tilde{\cal P}}=
(-1)^{m+1}\alpha_{\tilde{\cal P}}
 \quad , \quad
\inn ({\cal X}_{\alpha_{\tilde{\cal P}}})(\bar\kappa_{\tilde{\cal P}}^*\omega)=1
$$
(i.e., an extended HDW multivector field for $\hsmpio$) there exists
${\cal X}_{h_{\cal P}}\in\vf^m({\cal P})$ which 
is $h_{\cal P}$-related with ${\cal X}_\alpha$
and is a solution to the equations
$$
\inn ({\cal X}_{h_{\cal P}})\Omega_{h_{\cal P}}=0 \quad , \quad
\inn ({\cal X}_{h_{\cal P}})(\bar\tau_{\cal P}^*\omega)=1
$$
(i.e., a HDW multivector field for $\hsjpio$).
Furthermore, if ${\cal X}_{\alpha_{\tilde{\cal P}}}$ is integrable, then
${\cal X}_{h_{\cal P}}$ is integrable too.
\end{teor}

As a consequence, the following definition can be established:

\begin{definition}
Given an extended almost-regular global Hamiltonian system $\hsmpio$,
and considering all the Hamiltonian sections
 $h_{\cal P}\colon{\cal P}\to{\cal M}\pi$ such that
 ${\rm Im}\,h_{\cal P}$ are integral submanifolds of 
 ${\cal D}_{\alpha_{\cal P}}$,
 we have a family $\{ (J^1\pi^*,{\cal P},h_{\cal P})\}_{\alpha_{\tilde{\cal P}}}$, 
 which will be called the
 {\rm class of restricted almost-regular
 Hamiltonian systems associated with $\hsmpio$}.
\end{definition}

\begin{remark}
{\rm Observe that, for every Hamiltonian section $h_{\cal P}$ in this class,
${\rm Im}\,h_{\cal P}$ is a submanifold of $\tilde{\cal P}$.
Therefore we have induced a Hamiltonian section
$h_{\tilde{\cal P}}\colon{\cal P}\to\tilde{\cal P}$ of 
$\tilde\mu_{\tilde{\cal P}}$ such that
$h_{\cal P}=\jmath_{\tilde{\cal P}}\circ h_{\tilde{\cal P}}$.}
\label{pptilde}
\end{remark}

\begin{prop}
Let $\hsjpio$ be a restricted almost-regular Hamiltonian system.
\ben
\item
There exits a unique submanifold $\tilde{\cal P}$ of ${\cal M}\pi$ 
satisfying the conditions of Definition \ref{arehsdef},
and such that
$\mu_{\tilde{\cal P}}(\tilde{\cal P})={\cal P}$.
\item
Consider the submanifold $S={\rm Im}\, h_{\cal P}$
and the natural embedding
$\tilde\jmath_S\colon S={\rm Im}\, h_{\cal P}\hookrightarrow\tilde{\cal P}$.
 Then, there exists a unique local form
$\alpha_{\tilde{\cal P}}\in\df^1(\tilde{\cal P})$ such that:
\ben
\item
$\alpha_{\tilde{\cal P}}\in\ Z^1(\tilde{\cal P})$ (it is a closed form).
\item
$\tilde\jmath_S^*\alpha_{\tilde{\cal P}}=0$.
\item
$\inn(Y_{\tilde{\cal P}})\alpha_{\tilde{\cal P}}\not=0$, 
for every non-vanishing
$Y_{\tilde{\cal P}}\in\vf^{{\rm V}(\mu_{\tilde{\cal P}})}(\tilde{\cal P})$ and,
 in particular, such that
 \(\displaystyle\inn\left(\derpar{}{p}\right)\alpha_{\tilde{\cal P}}=1\),
 for every system of natural coordinates
 in $\tilde{\cal P}$, adapted to the bundle 
$\pi\colon E\to M$ (with $\omega=\d^mx$).
 \een
\een
\label{existuniq2}
\end{prop}
\proof
The existence and uniqueness of the submanifold $\tilde{\cal P}$
is assured, since it is made of all the fibers of $\mu$ at every
point of ${\cal P}$. 
The rest of the proof is like in Proposition \ref{existuniq}.
\qed

So we have the diagram
$$
\begin{array}{ccccc}
\tilde{\cal P} &
\begin{picture}(10,20)(0,0)
\put(-45,3){\vector(1,0){230}}
\put(65,8){\mbox{$\jmath_{\tilde{\cal P}}$}}
\end{picture}
 & & & {\cal M}\pi
\\
\begin{picture}(10,100)(0,0)
\put(5,100){\vector(0,-1){100}}
\put(-15,48){\mbox{$\tilde\mu_{\tilde{\cal P}}$}}
\end{picture}
&
\begin{picture}(100,100)(0,0)
\put(100,0){\vector(-1,1){100}}
\put(25,45){\mbox{$h_{\tilde{\cal P}}$}}
\end{picture}
&
\begin{picture}(10,100)(0,0)
\put(0,48){\mbox{$S$}}
\put(5,0){\vector(0,1){42}}
\put(15,55){\vector(2,1){95}}
\put(50,85){\mbox{$\jmath_S$}}
\put(-5,55){\vector(-2,1){95}}
\put(-50,85){\mbox{$\tilde\jmath_S$}}
\end{picture}
 &
\begin{picture}(100,100)(0,0)
\put(0,0){\vector(1,1){100}}
\put(60,45){\mbox{$h_{\cal P}$}}
\end{picture}
&
\begin{picture}(10,100)(0,0)
\put(5,100){\vector(0,-1){100}}
\put(10,48){\mbox{$\mu$}}
\end{picture}
\\
{\cal P} & 
\begin{picture}(100,10)(0,0)
\put(0,3){\vector(1,0){100}}
\put(45,6){\mbox{${\rm Id}$}}
\end{picture}
& {\cal P} &
\begin{picture}(100,10)(0,0)
\put(0,3){\vector(1,0){100}}
\put(45,9){\mbox{$\jmath_{\cal P}$}}
\end{picture}
& J^1\pi^* 
 \end{array}
 $$

Bearing in mind Remark \ref{pptilde}, we can also state:

\begin{corol}
Let ($\tilde{\cal P},\alpha_{\tilde{\cal P}})$ be
the couple associated with a 
given restricted almost-regular Hamiltonian system $\hsjpio$
by the Proposition \ref{existuniq2}.
Consider the characteristic distribution ${\cal D}_{\alpha_{\tilde{\cal P}}}$,
and let $\{h\}_{\alpha_{\tilde{\cal P}}}$ be the family of local sections of 
$\tilde\mu_{\tilde{\cal P}}$ such that 
${\rm Im}\,h\equiv S$ 
are local integral submanifolds of ${\cal D}_{\alpha_{\tilde{\cal P}}}$.
Then, for every 
$\tilde h'_{\cal P}\in\{ h\}_{\alpha_{\tilde{\cal P}}}$, 
we have that ${\rm Im}\,\tilde h'_{\cal P}$ is locally a
level set of a function ${\rm H}_{\tilde{\cal P}}$
such that 
$\alpha_{\tilde{\cal P}}\vert_S=(\d{\rm H}_{\tilde{\cal P}})\vert_S$, locally.
\end{corol}

\begin{definition}
Given a restricted almost-regular Hamiltonian system $\hsjpio$,
let ($\tilde{\cal P},\alpha_{\tilde{\cal P}})$ be
the couple associated with $\hsjpio$
by the Proposition \ref{existuniq2}.
The triple $({\cal M}\pi,\tilde{\cal P},\alpha_{\tilde{\cal P}})$
will be called the (local)
{\rm extended almost-regular Hamiltonian system associated with $\hsjpi$}.
\end{definition}

\begin{prop}
\ Let $\{ (J^1\pi^*,{\cal P},h_{\cal P})\}$ be the class of 
restricted almost-regular Hamiltonian systems 
associated with an extended almost-regular Hamiltonian system $\hsmpio$.
Consider the submanifolds $\{ S_{h_{\cal P}}={\rm Im}\, h_{\cal P}\}$,
for every Hamiltonian section $h$ in this class,
and let 
$\jmath_{S_{h_{\cal P}}}\colon S_{h_{\cal P}}\hookrightarrow\tilde{\cal P}$ be
the natural embeddings.
Then the submanifolds $\{ S={\rm Im}\, h_{\cal P}\}$,
for every Hamiltonian section $h_{\cal P}$ in this class
are premultisymplectomorphic.
\end{prop}

As a consequence of this,
if ${\cal X}_{\alpha_{\tilde{\cal P}}}\in\vf^m(\tilde{\cal P})$ is a solution
to the equations (\ref{hameq20}), the multivector fields
 ${\cal X}_{S_{h_{\cal P}}}\in\vf^m(S_{h_{\cal P}})$ such that 
$\Lambda^m(\tilde\jmath_{S_{h_{\cal P}}})_*{\cal X}_{S_{h_{\cal P}}}=
{\cal X}_{\alpha_{\tilde{\cal P}}}\vert_{S_{h_{\cal P}}}$,
for every submanifold $S_{h_{\cal P}}$ of this family,
are related by these presymplectomorphisms.

 \section{Examples}
\protect\label{exs}

\subsection{Restricted Hamiltonian system associated with a Lagrangian system}

A particular but relevant case concerns
(first-order) Lagrangian field theories and their
Hamiltonian counterparts.

In field theory, a {\sl Lagrangian system} is a couple $\ls$,
where $J^1\pi$ is the first-order jet bundle of $\pi\colon E\to M$, 
and $\Omega_\Lag\in\df^{m+1}(J^1\pi)$ is the
{\sl Poincar\'e-Cartan $(m+1)$-form}
associated with the Lagrangian density $\Lag$ describing the system
($\Lag$ is a $\bar\pi^1$-semibasic $m$-form on $J^1\pi$, 
which is written as
$\Lag =\lag \bar\pi^{1*}\eta\equiv\lag\omega$,
where $\lag\in\Cinfty (J^1\pi)$
is the {\sl Lagrangian function} associated with $\Lag$ and $\omega$).
The Lagrangian system is {\sl regular} if
$\Omega_{\Lag}$ is $1$-nondegenerate; elsewhere it is {\sl singular}.

The {\sl extended Legendre map} associated with $\Lag$,
 $\widetilde{{\cal F}\Lag}\colon J^1\pi\to {\cal M}\pi$,
 is defined by
  $$
 (\widetilde{{\cal F}\Lag}(\bar y))(\moment{Z}{1}{m}):=
 (\Theta_{\Lag})_{\bar y}(\moment{\bar Z}{1}{m})
 $$
 where $\moment{Z}{1}{m}\in\Tan_{\pi^1(\bar y)}E$, and
 $\moment{\bar Z}{1}{m}\in\Tan_{\bar y}J^1\pi$ are such that
 $\Tan_{\bar y}\pi^1\bar Z_{\alpha_{\tilde{\cal P}}}=Z_{\alpha_{\tilde{\cal P}}}$.
 ($\widetilde{{\cal F}\Lag}$ can also be defined as the
  ``first order  vertical Taylor approximation to
 $\lag$'' \cite{CCI-91}).
 We have that $\widetilde{{\cal F}\Lag}^*\Omega=\Omega_{\Lag}$.
If $(x^\alpha,y^A,v^A_\alpha)$ is a natural chart of coordinates in $J^1\pi$
(adapted to the bundle structure, and such that
$\omega=\d x^1\wedge\ldots\wedge\d x^m\equiv\d x^m$)
 the local expressions of these maps are
 $$
 \begin{array}{ccccccc}
 \widetilde{{\cal F}\Lag}^*x^\nu = x^\nu &\quad\ , \ \quad&
 \widetilde{{\cal F}\Lag}^*y^A = y^A &\quad\  , \quad&
 \widetilde{{\cal F}\Lag}^*p_A^\nu =\derpar{\lag}{v^A_\nu}
 &\quad\ , \quad&
 \widetilde{{\cal F}\Lag}^*p =\lag-v^A_\nu\derpar{\lag}{v^A_\nu}
 \\
 {\cal F}\Lag^*x^\nu = x^\nu &\quad\ , \ \quad&
 {\cal F}\Lag^*y^A = y^A &\quad\  , \quad&
 {\cal F}\Lag^*p_A^\nu =\derpar{\lag}{v^A_\nu} & &
 \end{array}
 $$

 Using the natural projection
 $\mu \colon {\cal M}\pi\to J^1\pi^*$,
 we define the {\sl restricted Legendre map} associated with $\Lag$ as
 ${\cal F}\Lag :=\mu\circ\widetilde{{\cal F}\Lag}$.

 Then, $\ls$ is a {\sl regular}
 Lagrangian system if ${\cal F}\Lag$ is a local diffeomorphism
 (this definition is equivalent to that given above).
 Elsewhere $\ls$ is a {\sl singular} Lagrangian system.
 As a particular case, $\ls$ is a {\sl hyper-regular}
 Lagrangian system if ${\cal F}\Lag$ is a global diffeomorphism.
 Finally, a singular Lagrangian system $\ls$ is {\sl almost-regular} if:
  ${\cal P}:={\cal F}\Lag (J^1\pi)$ is a closed submanifold of $J^1\pi^*$,
  ${\cal F}\Lag$ is a submersion onto its image, and
  for every $\bar y\in J^1\pi$, the fibers
  ${\cal F}\Lag^{-1}({\cal F}\Lag (\bar y))$
  are connected submanifolds of $J^1E$.

If $\ls$ is a hyper-regular Lagrangian system, then
 $\widetilde{{\cal F}\Lag}(J^1\pi)$ is a
 1-codimensional imbedded submanifold of ${\cal M}\pi$,
 which is transverse to the projection $\mu$, and is diffeomorphic to
 $J^1\pi^*$. This diffeomorphism is $\mu^{-1}$, when $\mu$ is
 restricted to $\widetilde{{\cal F}\Lag}(J^1\pi)$, and also coincides with the map
 $h:=\widetilde{{\cal F}\Lag}\circ{\cal F}\Lag^{-1}$,
 when it is restricted onto its image
 (which is just $\widetilde{{\cal F}\Lag}(J^1\pi)$).
 This map $h$ is the Hamiltonian section needed
 to construct the {\sl restricted Hamiltonian system associated with $\ls$}.
 In other words, the Hamiltonian section $h$ is given by the image
 of the extended Legendre map.

 Using charts of natural coordinates in $J^1\pi^*$ and
 ${\cal M}\pi$, we obtain that the local Hamiltonian function
 ${\rm h}$ representing this Hamiltonian section is
 $$
 {\rm h}(x^\nu,y^A,p^\nu_A)=
 ({\cal F}\Lag^{-1})^*\left(v^A_\nu\derpar{\lag}{v^A_\nu}-\lag\right)=
 p^\nu_A({\cal F}\Lag^{-1})^*v_\nu^A- {\cal F}\Lag^{-1^*}\lag
 $$
Of course, if $\hsmpi$ is any extended Hamiltonian system associated with
$\hsjpi$, then $\widetilde{{\cal F}\Lag}(J^1\pi)$ is an integral
submanifold of the characteristic distribution of $\alpha$.

 In an analogous way, if $\ls$ is an almost-regular Lagrangian system,
 and the submanifold 
 $\jmath_{\tilde{\cal P}}\colon{\cal P}\hookrightarrow J^1\pi^*$
 is a fiber bundle over $E$ and $M$,
 the $\mu$-transverse submanifold
 $\jmath\colon\widetilde{{\cal F}\Lag}(J^1\pi)\hookrightarrow{\cal M}\pi$
 is diffeomorphic to ${\cal P}$. This diffeomorphism
 $\tilde\mu\colon\widetilde{{\cal F}\Lag}(J^1\pi)\to{\cal P}$
 is just the restriction of the projection $\mu$ to
 $\widetilde{{\cal F}\Lag}(J^1\pi)$.
 Then, taking the Hamiltonian section $h_{\cal P}:=\jmath\circ\tilde\mu^{-1}$,
 we define the Hamilton-Cartan forms
  $$
 \Theta_{h_{\cal P}}=h_{\cal P}^*\Theta
 \quad  \quad
 \Omega_{h_{\cal P}}=h_{\cal P}^*\Omega
 $$
 which verify that
 ${\cal F}\Lag_{\cal P}^*\Theta^0_h=\Theta_{\Lag}$ and
 ${\cal F}\Lag_{\cal P}^*\Omega^0_h=\Omega_{\Lag}$
(where ${\cal F}\Lag_{\cal P}$ is the restriction map of
${\cal F}\Lag$ onto ${\cal P}$).
 Once again, this Hamiltonian section $h_{\cal P}$ is given by the image
 of the extended Legendre map.
 Then $\hsjpio$ is the {\sl Hamiltonian system}
 associated with the almost-regular Lagrangian system $\ls$,
 and we have the following diagram
 $$
\begin{array}{cccc}
\begin{picture}(15,60)(0,0)
\put(0,0){\mbox{$J^1\pi$}}
\end{picture}
&
\begin{picture}(65,52)(0,0)
 \put(15,28){\mbox{$\widetilde{{\cal F}\Lag}_{\cal P}$}}
 \put(24,-9){\mbox{${\cal F}\Lag_{\cal P}$}}
 \put(0,7){\vector(2,1){60}}
 \put(0,2){\vector(1,0){65}}
\end{picture}
&
\begin{picture}(90,52)(0,0)
 \put(5,0){\mbox{${\cal P}$}}
 \put(-15,42){\mbox{$\widetilde{{\cal F}\Lag}(J^1\pi)$}}
 \put(5,13){\vector(0,1){25}}
 \put(10,38){\vector(0,-1){25}}
 \put(-15,20){\mbox{$\tilde\mu^{-1}$}}
 \put(12,22){\mbox{$\tilde\mu$}}
 \put(35,45){\vector(1,0){55}}
 \put(30,2){\vector(1,0){55}}
 \put(30,8){\vector(2,1){55}}
 \put(53,-9){\mbox{$\jmath_{\cal P}$}}
 \put(48,33){\mbox{$\jmath$}}
 \put(65,12){\mbox{$h_{\cal P}$}}
 \end{picture}
&
\begin{picture}(15,52)(0,0)
 \put(0,0){\mbox{$J^1\pi^*$}}
 \put(0,41){\mbox{${\cal M}\pi$}}
 \put(10,38){\vector(0,-1){25}}
 \put(0,22){\mbox{$\mu$}}
\end{picture}
\\
& &
\begin{picture}(90,35)(0,0)
 \put(10,35){\vector(1,-1){35}}
 \put(5,11){\mbox{$\bar\tau_{\cal P}$}}
 \put(90,11){\mbox{$\bar\tau$}}
 \put(100,35){\vector(-1,-1){35}}
\end{picture}
 &
\\
& & \quad\quad M &
 \end{array}
 $$

The construction of the (local) extended almost-regular Hamiltonian system
associated with $\hsjpio$ can be made by following the procedure
described in section \ref{rearrhs}.
Of course, if $\hsmpio$ is the extended Hamiltonian system associated with
$\hsjpio$, then $\widetilde{{\cal F}\Lag}(J^1\pi)$ is an integral
submanifold of the characteristic distribution of $\alpha_{\tilde{\cal P}}$.

\subsection{Non-autonomous dynamical systems}
\protect\label{nads}

Another example consists in showing how the
so-called {\sl extended formalism of time-dependent mechanics}
(see \cite{GMS-97}, \cite{Ku-tdms}, \cite{MS-98}, \cite{Ra1}, \cite{St-2005})
can be recovered from this more general framework.

The starting point consists in giving the configuration bundle, which
for a large class of non-autonomous dynamical systems can be taken
to be $\pi\colon E\equiv Q\times\Real\to\Real$, where $Q$ is a
$n$-dimensional differentiable manifold endowed with local
coordinates $(q^i)$, and $\Real$ has as a global coordinate $t$.
The {\sl extended} and {\sl restricted momentum phase spaces} are
$$ {\cal M}\pi \simeq \Tan^*E\equiv \Tan^*(Q\times\Real )
 \simeq \Tan^*Q \times \Real \times \Real^*
\quad , \quad
J^1\pi^* \simeq \Tan^*Q\times\Real
$$
Then, the following projections can be defined
\beann
pr_1 \colon{\cal M}\pi \to \Tan^*Q
&,&
\mu \colon{\cal M}\pi \to \Tan^*Q\times\Real
\\
pr_2 \colon{\cal M}\pi \to \Real \times \Real^*
&,&
p \colon{\cal M}\pi \to \Real^*
\eeann
If $\Omega_Q \in Z^2(\Tan^*Q)$ and
$\Omega_{\Real} \in Z^2(\Real \times \Real^*)$
denote the natural symplectic forms of
$\Tan^*Q$ and $\Real \times \Real^*$, then
the natural symplectic structure of ${\cal M}\pi$ is just
$$
\Omega = pr_1^* \Omega_Q + pr_2^* \Omega_{\Real}.
$$
Then, we define the so-called
{\sl extended time-dependent Hamiltonian function}
$$
{\rm H} := \mu^*{\rm h}+p\in \Cinfty(\Tan^*(Q\times\Real))
$$
where the dynamical information is given
by the ``time-dependent Hamiltonian function''
${\rm h}\in\Cinfty(\Tan^*Q\times\Real)$.

Now we have that $(\Tan^*(Q\times\Real),\Omega,\alpha)$,
with $\alpha=\d{\rm H}$, is an
{\sl extended global Hamiltonian system}, and then the equations of
motion are 
\beq 
\inn(X_{\rm H})\Omega = \d{\rm H} \quad , \quad
\inn(X_{\rm H})\d t = 1 \qquad  
\mbox{\rm with $X_{\rm H} \in\vf(\Tan^*(Q\times\Real)) $}
\label{ecs} 
\eeq 
In order to analyze the
information given by this equation, we take a local chart of
coordinates $(q^i,p_i,t,p)$ in $\Tan^*(Q\times\Real)$, and one can
check that the unique solution to these equations is 
\bea
 X_{\rm H} &=& \derpar{{\rm H}}{p_i} \derpar{}{q^i} - 
\derpar{{\rm H}}{q^i} \derpar{}{p_i} + \derpar{}{t} - 
\derpar{{\rm H}}{t}\derpar{}{p} \nonumber
 \\ &=& 
\derpar{(\mu^*{\rm h})}{p_i}
\derpar{}{q^i} - \derpar{(\mu^*{\rm h})}{q^i} \derpar{}{p_i} +
\derpar{}{t} - \derpar{(\mu^*{\rm h})}{t}\derpar{}{p}
 \label{XH}
\eea 
If $\tilde\psi(t)=(q^i(t),p_i(t),t,p(t))$ denote the integral
curves of this vector field, the last expression leads to the
following system of extended Hamiltonian equations
 \beq
\frac{d(q^i\circ\tilde\psi)}{dt}=\derpar{(\mu^*{\rm h})}{p_i}\circ\tilde\psi
 \quad ,\quad
\frac{d(p_i\circ\tilde\psi)}{dt}=-\derpar{(\mu^*{\rm h})}{q^i}\circ\tilde\psi
 \quad ,\quad
\frac{d(p\circ\tilde\psi)}{dt}=-\derpar{(\mu^*{\rm h})}{t}\circ\tilde\psi
\label{eqfinal}
 \eeq
Observe that the last equation corresponds to the
last group of equations (\ref{ehdwe}) in the general case of field theories.
In fact, using the other Hamilton equations we get
\beann
\frac{d(p\circ\tilde\psi)}{dt}&=&-\frac{(\mu^*{\rm h}\circ\psi)}{dt}=
-\derpar{(\mu^*{\rm h})}{t}\Big\vert_{\tilde\psi}-
\derpar{(\mu^*{\rm h})}{q^i}\Big\vert_{\tilde\psi}\frac{(q^i\circ\tilde\psi)}{t}-
\derpar{(\mu^*{\rm h})}{p_i}\Big\vert_{\tilde\psi}\frac{(p_i\circ\tilde\psi)}{t}
\\ &=&
-\derpar{(\mu^*{\rm h})}{t}\Big\vert_{\tilde\psi}-
\derpar{(\mu^*{\rm h})}{q^i}\Big\vert_{\tilde\psi}
\derpar{(\mu^*{\rm h})}{p_i}\Big\vert_{\tilde\psi}+
\derpar{(\mu^*{\rm h})}{p_i}\Big\vert_{\tilde\psi}
\derpar{(\mu^*{\rm h})}{q^i}\Big\vert_{\tilde\psi}=
-\derpar{(\mu^*{\rm h})}{q^i}\Big\vert_{\tilde\psi}
\eeann

However, as the physical states are the points of $\Tan^*Q\times\Real$
and not those of $\Tan^*(Q\times\Real)$, the vector field
which gives the real dynamical evolution is not
$X_{\rm H}$, but another one in $\Tan^*Q\times\Real$
which, as $X_{\rm H}$ is $\mu$-projectable, is just
$\mu_* X_{\rm H}=X_h \in\vf(\Tan^*Q\times\Real)$, that is,
in local coordinates $(q^i,p_i,t)$ of $\Tan^*Q\times\Real$,
\beq
X_h =
\derpar{{\rm h}}{p_i}\derpar{}{q^i}-\derpar{{\rm h}}{q^i}\derpar{}{p_i} + 
\derpar{}{t}
\label{Xh}
\eeq
Thus, the integral curves $\psi(t)=(q^i(t),p_i(t),t)$ of $X_h$
are the $\mu$-projection of those of $X_H$, and they
are solutions to the system of Hamilton equations
$$
\frac{d(q^i\circ\psi)}{dt} = \derpar{h}{p_i}\circ\psi \quad ,\quad
\frac{d(p_i\circ\psi)}{dt} = - \derpar{h}{q^i}\circ\psi 
$$

This result can also be obtained by considering the class of
restricted Hamiltonian systems associated with
$(\Tan^*(Q\times\Real),\Omega,\d{\rm H})$.
In fact, $\Tan^*(Q\times\Real)$
is foliated by the family of hypersurfaces of $\Tan^*(Q\times\Real)$
where the extended Hamiltonian function is constant; that is,
$$
S := \{{\bf p}\in\Tan^*(Q\times\Real) \ \mid \ {\rm H}({\bf p})= r\ (ctn.) \}
$$
which are the integral submanifolds of the characteristic distribution
of $\alpha=\d{\rm H}$.
Thus, every $S$ is defined in $\Tan^*(Q\times\Real)$
by the constraint $\zeta := {\rm H}-r$, and the vector field
given in (\ref{XH}), which is the solution to (\ref{ecs}),
is tangent to all of these submanifolds.
Then, taking the global Hamiltonian sections
$$
h\colon (q^i,p_i,t) \mapsto (q^i,p_i,t, p=r-\mu^*{\rm h})
$$
we can construct the restricted Hamiltonian systems
$(\Tan^*Q\times\Real,h)$ associated with
$(\Tan^*(Q\times\Real),\Omega,\d{\rm H})$. Therefore
(\ref{Xh}) is the solution to the equations
$$
\inn(X_h)\Omega_h = 0 \quad , \quad \inn(X_h)\d t = 1
\qquad \mbox{\rm with $X_{\rm H} \in \vf(\Tan^*(Q\times\Real))$}
$$
where
$$
\Omega_h=h^*\Omega=\omega_Q -\d{\rm h}\wedge\d t\in\df^2(\Tan^*Q\times\Real)
$$
The dynamics on each one of these
restricted Hamiltonian systems is associated
to a given constant value of the extended Hamiltonian.
Observe also that, on every submanifol $S$,
the global coordinate $p$ is identified with the physical energy 
by means of the time-dependent Hamiltonian function $\mu^*{\rm h}$, 
and hence the last equation (\ref{eqfinal}) shows the known fact that the
energy is not conserved on the dynamical trajectories
of time-dependent systems.

In this way, we have also recovered one of the standard
Hamiltonian formalisms for time-dependent systems
(see \cite{EMR-91}).

 \section{Conclusions and outlook}

The usual way of defining Hamiltonian systems in first-order field theory
consists in working in the restricted multimomentum bundle
$J^1\pi^*$, which is the natural multimomentum phase space for field theories,
but $J^1\pi^*$ has no a natural multisymplectic structure.
Thus, in order to define restricted Hamiltonian systems 
we use Hamiltonian sections $h\colon J^1\pi^*\to{\cal M}\pi$,
which carry the `physical information' and allow us to pull-back
the natural multisymplectic structure of ${\cal M}\pi$ to $J^1\pi^*$.
In this way we obtain the Hamilton-Cartan form
$\Omega_h\in\df^{m+1}(J^1\pi^*)$, 
and then the Hamiltonian field equations can be derived from the 
Hamilton-Jacobi variational principle. As a consequence,
both the geometry and the `physical information' are coupled
in the non-canonical multisymplectic form $\Omega_h$.

The alternative way that we have introduced consists in working 
directly in the extended multimomentum bundle 
${\cal M}\pi$,
which is endowed with a canonical multisymplectic structure 
$\Omega\in\df^{m+1}({\cal M}\pi)$.
Then we define extended Hamiltonian systems as
a triple $\hsmpi$, where $\alpha\in Z^1({\cal M\pi})$
is a $\mu$-transverse closed form, and the Hamiltonian equation
is $\inn({\cal X})\Omega=(-1)^{m+1}\alpha$, with ${\cal X}\in\vf^m({\cal M}\pi)$.
Thus, in these models, 
the geometry $\Omega$ and the `physical information' $\alpha$
are not coupled, and geometric field equations
can be expressed in an analogous way to those 
of mechanical autonomous Hamiltonian systems.

The characteristic distribution ${\cal D}_\alpha$ associated with
$\alpha$, being involutive,
has 1-codimensional and $\mu$-transverse 
integrable submanifolds of ${\cal M}\pi$,
where the sections solution to the field equations are contained.
These integrable submanifolds can be locally identified 
with local sections of the affine bundle
$\mu\colon{\cal M}\pi\to J^1\pi^*$.
Each one of them allows us to define locally a restricted Hamiltonian system,
although all those associated with the same form $\alpha$
are, in fact, multisymplectomorphic.
The conditions for the existence of global Hamiltonian sections
have been also analyzed.
Conversely, every restricted
Hamiltonian system is associated with an extended
Hamiltonian system (at least locally).

In addition, the extended Hamiltonian field equations can be obtained from an
{\sl extended Hamilton-Jacobi variational principle},
stated on the set of sections of the bundle
$\bar\kappa\colon{\cal M}\pi\to M$, which are integral sections of 
the characteristic distribution of $\alpha$, taking the variations
given by the set of the $\bar\kappa$-vertical vector fields incident
to $\alpha$.
In fact, a part of the local system of differential equations 
for the critical sections of an extended Hamiltonian system
is the same as for the associated restricted Hamiltonian system.
Nevertheless, there is another part of the whole system 
of differential equations which leads to the condition
that the critical sections must also be integral submanifolds
of the characteristic distribution ${\cal D}_\alpha$.

Restricted and extended Hamiltonian systems
for submanifolds of $J^1\pi^*$ and ${\cal M}\pi$ 
(satisfying suitable conditions)
have been defined in order to include the almost-regular
field theories in this picture.
Their properties are analogous to the former case.

The extended Hamiltonian formalism has already been used
for defining Poisson brackets in field theories \cite{FPR-2003}.
It could provide new insights into some classical problems,
such as: reduction of multisymplectic Hamiltonian systems
with symmetry, integrability, and quantization of multisymplectic
Hamiltonian field theories.

\subsection*{Acknowledgments}

We acknowledge the financial support of
{\sl Ministerio de Educaci\'on y Ciencia}, projects
 BFM2002-03493, MTM2004-7832 and MTM2005-04947.
We wish to thank Mr. Jeff Palmer for his
assistance in preparing the English version of the manuscript.


\begin{thebibliography}{MM}

{\small

\bibitem{AA-80}
{\sc V. Aldaya, J.A. de Azc\'arraga},
``Geometric formulation of classical mechanics and field theory'',
{\sl Riv. Nuovo Cimento} {\bf 3}(10) (1980) 1-66.

\bibitem{Aw-92}
{\sc A. Awane}, ``$k$-symplectic structures'', {\sl J. Math.
Phys.} {\bf 32}(12) (1992) 4046-4052.

\bibitem{BSF-88}
{\sc E. Binz, J. Sniatycki, H. Fisher}, {\sl The Geometry of
Classical fields}, North Holland, Amsterdam, 1988.

\bibitem{Ca-96a} 
{\sc F. Cantrijn, A. Ibort, M. de Le\'on},
``Hamiltonian Structures on Multisymplectic
manifolds''.  {\sl Rend. Sem. Mat. Univ. Pol. Torino} {\bf 54},
 (1996) 225-236.

\bibitem{CIL-98}
{\sc F. Cantrijn, L.A. Ibort, M. de Le\'on}, ``On the Geometry of
Multisymplectic Manifolds'', {\sl J. Austral. Math. Soc. Ser.}
{\bf 66} (1999) 303-330.

\bibitem{CCI-91}
{\sc J.F. Cari\~nena, M. Crampin, L.A. Ibort}, ``On the
multisymplectic formalism for first order Field Theories'',
 {\sl Diff. Geom. Appl.} {\bf 1} (1991) 345-374.

\bibitem{CM-2003}
{\sc M. Castrill\'on-L\'opez, J.E. Marsden},
``Some remarks on Lagrangian and Poisson reduction for
field theories'',
{\sl J. Geom. Phys.} {\bf 48}(1) (2003) 52-83.

\bibitem{EMR-91}
{\sc A. Echeverr\'\i a-Enr\'\i quez, M.C. Mu\~noz-Lecanda, N.
Rom\'an-Roy}, ``Geometrical setting of time-dependent regular
systems. Alternative models'', {\sl Rev. Math. Phys.} {\bf 3}(3)
(1991) 301-330.

\bibitem{EMR-96}
{\sc A. Echeverr\'\i a-Enr\'\i quez, M.C. Mu\~noz-Lecanda, N. Rom\'an-Roy},
``Geometry of Lagrangian first-order classical field theories''.
{\sl Forts. Phys.} {\bf 44} (1996) 235-280.

\bibitem{EMR-98}
{\sc A. Echeverr\'\i a-Enr\'\i quez, M.C. Mu\~noz-Lecanda, N. Rom\'an-Roy},
``Multivector Fields and Connections.
Setting Lagrangian Equations in Field Theories''.
{\sl J. Math. Phys.} {\bf 39}(9) (1998) 4578-4603.

\bibitem{EMR-99b}
{\sc A. Echeverr\'\i a-Enr\'\i quez, M.C. Mu\~noz-Lecanda, N.
 Rom\'an-Roy}, ``Multivector Field Formulation of Hamiltonian
 Field Theories: Equations and Symmetries'',
 {\sl J. Phys. A: Math. Gen.} {\bf 32} (1999) 8461-8484.

\bibitem{EMR-00}
{\sc A. Echeverr\'\i a-Enr\'\i quez, M.C. Mu\~noz-Lecanda, N.
 Rom\'an-Roy}, ``Geometry of Multisymplectic Hamiltonian First-order
 Field Theories'', {\sl J. Math. Phys.} {\bf 41}(11) (2000) 7402-7444.

\bibitem{EMR-2002}
{\sc A. Echeverr\'\i a-Enr\'\i quez, M.C. Mu\~noz-Lecanda, N.
 Rom\'an-Roy}, ``A Geometrical Analysis of the Field equations
 in Field Theories'',
 {\sl Int. J. Math. Math. Sci.} {\bf 29}(12) (2002) 687-699.

\bibitem{FPR-2003}
{\sc M. Forger, C. Paufler, H. R\"omer},
``A general construction of Poisson brackets on
exact multisymplectic manifolds'', 
{\sl Rep. Math. Phys.} {\bf 51}(2-3) (2003) 187-195.

\bibitem{Gc-73}
{\sc P.L. Garc\'ia}, ``The Poincar\'e-Cartan invariant in the
calculus of variations'', {\sl Symp. Math.} {\bf 14} (Convegno di
Geometria Simplettica e Fisica Matematica, INDAM, Rome, 1973),
Acad. Press, London  (1974) 219-246.

\bibitem{GMS-97b}
{\sc G. Giachetta, L. Mangiarotti, G. Sardanashvily}, {\sl New
Lagrangian and Hamiltonian Methods in Field Theory}, World
Scientific Pub. Co., Singapore (1997).

\bibitem{GMS-97}
{\sc G. Giachetta, L. Mangiarotti, G. Sardanashvily},
``Differential Geometry of Time-Dependent Mechanics'',
dg-ga/9702020.

\bibitem{GS-73}
{\sc H. Goldschmidt, S. Sternberg}, ``The Hamilton-Cartan
formalism in the calculus of variations'', {\sl Ann. Inst. Fourier
Grenoble} {\bf 23}(1) (1973) 203-267.

\bibitem{Go-91b}
{\sc M.J. Gotay}, ``A Multisymplectic Framework for Classical
Field Theory and the Calculus of Variations I: Covariant
Hamiltonian formalism'', {\sl Mechanics, Analysis and Geometry:
200 Years after Lagrange}, M. Francaviglia Ed., Elsevier Science
Pub. (1991) 203-235.

\bibitem{GIMMSY-mm}
{\sc M.J. Gotay, J.Isenberg, J.E. Marsden, R. Montgomery},
{\sl Momentum maps and classical relativistic fields I: Covariant
Theory}, physics/9801019 (v2), (2003).

\bibitem{Gu-87}
{\sc C G\"unther}, ``The polysymplectic Hamiltonian formalism in
the Field Theory and the calculus of variations I: the local
case'', {\sl J. Diff. Geom.} {\bf 25} (1987) 23-53.

\bibitem{HK-01}
{\sc F. H\'elein, J. Kouneiher},
``Finite dimensional Hamiltonian formalism for gauge and quantum field theories'',
{\sl J. Math. Phys.} {\bf 43}(5) (2002) 2306-2347.

\bibitem{Ka-98}
{\sc I.V. Kanatchikov},
``Canonical structure of Classical Field Theory in the polymomentum
phase space'',
{\sl Rep. Math. Phys.} {\bf 41}(1) (1998) 49-90.

\bibitem{KS-75}
{\sc J. Kijowski, W. Szczyrba}, ``Multisymplectic Manifolds and
the Geometrical Construction of the Poisson Brackets in the
Classical Field Theory'', {\sl G\'eom\'etrie Symplectique et
Physique Math\'ematique} Coll. Int. C.N.R.S. {\bf 237} (1975)
347-378.

\bibitem{KT-79}
{\sc J. Kijowski, W.M. Tulczyjew}, {\sl A Symplectic Framework for
Field Theories}, Lect. Notes Phys. {\bf 170}, Springer-Verlag,
berlin (1979).

 \bibitem{Kr-87}
{\sc D. Krupka},
 ``Regular Lagrangians and Lepagean forms'',
 Proc. on {\sl Diff. Geom. Appls. (Brno 1986)},
Math Ap. (East European Ser. {\bf 27}), Reidel, Dordrecht (1987) 111-148.

\bibitem{KS-01a}
{\sc O. Krupkova, D. Smetanova},
``On regularization of variational problems in first-order field theory''.
 Proc. on {\sl $20th$ Winter School on Geom. and Phys.},
{\sl Rend. Circ. Mat. Palermo (2) Suppl.} {\bf 66} (2001) 133-140.

\bibitem{KS-01b}
{\sc O. Krupkova, D. Smetanova},
``Legendre transformation for regularizable Lagrangians in field theory'',
{\sl Lett. Math. Phys.} {\bf 58}(3) (2002) 189-204.

\bibitem{Ku-tdms}
{\sc R. Kuwabara},
``Time-dependent mechanical symmetries and extended Hamiltonian systems'',
{\sl Rep. Math. Phys.} {\bf 19} (1984) 27-38.

\bibitem{LMM-95}
{\sc M. de Le\'on, J. Mar\'\i n-Solano, J.C. Marrero}, ``Ehresmann
Connections in Classical Field Theories'', {\sl Proc. III Fall
Workshop: Differential Geometry and its Applications}, Anales de
F\'\i sica, Monograf\'\i as {\bf 2} (1995) 73-89.

 \bibitem{LMM-96}
{\sc M. de Le\'on, J. Mar\'\i n-Solano, J.C. Marrero},
``A Geometrical approach to Classical Field Theories: A constraint
algorithm for singular theories'',
Proc. on {\sl New Developments in Differential Geometry},
 L. Tamassi-J. Szenthe eds., Kluwer Acad. Press, (1996) 291-312.

 \bibitem{LMMMR-04}
{\sc M. de Le\'on, J. Mar\'\i n-Solano, J.C. Marrero, M.C. Mu\~noz-Lecanda,
 N. Rom\'an-Roy},
``Multisymplectic constraint algorithm for field theories'',
 {\sl Int. J. Geom. Meth. Mod. Phys.} {\bf 2}(5) (2005) 839-871.

\bibitem{LMS-2003}
{\sc M. de Le\'on, D. Mart\'\i n de Diego, A. Santamar\'\i a-Merino},
``Tulczyjew triples and Lagrangian submanifolds in classical field theories'',
in {\sl Applied Differential Geometry and Mechanics};
W. Sarlet and F. Cantrijn eds. Univ. of Gent, Gent, Academia Press (2003)
21-47.

\bibitem{LMS-2004}
{\sc M. de Le\'on, D. Mart\'\i n de Diego, A. Santamar\'\i a-Merino},
``Symmetries in classical field theories'',
{\sl Int. J. Geom. Meth. Mod. Phys.} {\bf 1}(5) (2004) 651-710.

\bibitem{mod2}
{\sc M. de Le\'{o}n, E. Merino,  M. Salgado},
``$k$-cosymplectic manifolds and Lagrangian field theories'',
{\sl J. Math. Phys.} {\bf 42}(5) (2001) 2092--2104.

\bibitem{MS-98}
{\sc L. Mangiarotti, G. Sardanashvily},
``Gauge Mechanics'', {\sl World Scientific}, Singapore, 1998.

\bibitem{MS-99}
{\sc J.E. Marsden, S. Shkoller}, ``Multisymplectic Geometry,
Covariant Hamiltonians and Water Waves'', {\sl Math. Proc. Camb.
Phil. Soc.} {\bf 125} (1999) 553-575.

\bibitem{Ma-2004}
{\sc E. Mart\'\i nez},
``Classical Field theory on Lie algebroids: multisymplectic formalism'',
math.DG/0411352 (2004).

\bibitem{Ma-2005}
{\sc E. Mart\'\i nez},
``Classical field theory on Lie algebroids: variational aspects'',
{\sl  J. Phys. A: Math. Gen.} {\bf 38}(32) (2005) 7145-7160

\bibitem{fam}
{\sc F. Munteanu, A.M. Rey, M. Salgado},
``The G\"{u}nther's formalism in classical field theory:
momentum map and reduction'',
{\sl J. Math. Phys.} {bf 45}(5) (2004) 1730--1751.

\bibitem{PR-2002}
{\sc C. Paufler, H. R\"omer},
``Geometry of Hamiltonian $n$-vector fields in multisymplectic
field theory'', {\sl J. Geom. Phys.} {\bf 44}(1) (2002) 52-69.

\bibitem{Ra1}
{\sc M. F. Ra\~nada},
 ``Extended Legendre transformation approach to the time-dependent Hamiltonian
formalism'', {\sl J. Phys. A: Math. Gen.} {\bf 25} (1992) 4025-4035.

\bibitem{Sd-95}
{\sc G. Sardanashvily},
{\sl Generalized Hamiltonian Formalism for Field Theory. Constraint Systems},
World Scientific, Singapore (1995).

\bibitem{Sa-89}
{\sc D.J. Saunders}, {\sl The Geometry of Jet Bundles}, London
Math. Soc. Lect. Notes Ser. {\bf 142}, Cambridge, Univ. Press,
1989.

\bibitem{So-69}
{\sc J.M. Souriau},
{\sl Structure des syst\'emes dynamiques\/},
Dunod, Paris, 1969.

\bibitem{St-2005}
{\sc J. Struckmeier},
``Hamiltonian dynamics on the symplectic extended phase space
for autonomous and non-autonomous systems'',
{\sl J. Phys. A: Math. Gen.} {\bf 38} (2005) 1275--1278.

\bibitem{VCLM-2005}
{\sc J. Vankerschaver, F. Cantrijn, M. de Le\'on, D. Mart\'\i n de Diego},
``Geometric aspects of nonholonomic field theories'',
{\sl Rep. Math. Phys.} {\bf 56}(3)  (2005) 387--411
}

\end{thebibliography}
\end{document}